\documentclass{article}
\usepackage{paper}

\begin{document}
\title{Surface Oxides Form on Pt(111)—But Vanish During Ammonia Oxidation}

\author[1,2]{D. Simonne\thanks{Corresponding author, dsimonne@mit.edu}}
\author[1]{A. Coati}
\author[1]{A. Vlad}
\author[1,4]{Y. Garreau}
\author[1]{B. Voisin}
\author[2,3]{M.I. Richard}
\author[1]{A. Resta}

\affil[1]{Synchrotron SOLEIL, L’Orme des Merisiers, Départementale 128, 91190 Saint-Aubin, France}
\affil[2]{Université Grenoble Alpes, CEA Grenoble, IRIG, MEM, NRX, 17 avenue des Martyrs, 38000 Grenoble, France}
\affil[3]{European Synchrotron Research Facility, 71 Avenue des Martyrs, 38000 Grenoble, France}
\affil[4]{Laboratoire Matériaux et Phénomènes Quantiques, Université Paris Cité, CNRS, 10 Rue Alice Domon et Léonie Duquet, 75013 Paris, France}
\maketitle

\begin{abstract}

Ammonia oxidation on platinum catalysts is pivotal for industrial nitric acid production and environmental abatement, yet the role of surface oxides in this process remains debated. 
Using \textit{operando} surface X-ray diffraction (SXRD), crystal truncation rod (CTR) analysis, and near-ambient pressure X-ray photoelectron spectroscopy (NAP-XPS), we reveal that Pt(111) does not form stable surface oxides under ammonia oxidation conditions. 
Instead, transient hexagonal monolayers and a Pt(111)-(8×8) superstructure emerge under oxygen-rich atmospheres and above the catalyst light-off temperature, but vanish upon ammonia exposure. 
Real-time mass spectrometry and NAP-XPS demonstrate that the reaction proceeds via a Langmuir-Hinshelwood mechanism, where adsorbed \ce{NH_x} and \ce{O} species availability dictate selectivity toward \ce{NO} or \ce{N2}.
Reducing the oxygen pressure by an order of magnitude slows the kinetics of oxide growth, only detected after \qty{24}{\hour}, and facilitated by transient and precursor structures.

\end{abstract}


\section{Introduction}

Ammonia oxidation on platinum catalysts is central to nitric acid production by the Ostwald process, where high selectivity toward nitric oxide (\ce{NO}) is required \parencite{Hatscher2008}. 
The difficulty is that \ce{NH3} oxidation proceeds through several competing routes that also generate \ce{N2O} and \ce{N2}. 
In practice, maximizing \ce{NO} formation requires high temperature, oxygen-rich feeds (high $p_{\ce{O2}}/p_{\ce{NH3}}$), and a surface that remains catalytically active under strongly oxidizing conditions \parencite{Heck1982, Hatscher2008}. 
Beyond nitric acid synthesis, this reaction is environmentally and technologically relevant. 
Selectivity control between \ce{NO} and \ce{N2} affects nitrogen-oxide emissions and enables routes for ammonia abatement \parencite{MITCLIMATE, EPAGreenhouseGases, Solomon2007, Davidson2009, Kabange2022}. 
In addition, renewed interest in ammonia as an energy carrier motivates a mechanistic understanding of its oxidation and decomposition chemistry at catalytic interfaces \parencite{Afif2016, Georgina2021}. 
From a surface-science standpoint, ammonia oxidation is also an attractive model reaction: it couples strong adsorbate interactions with oxidizing environments, and it drives structural and chemical transformations that are both facet-dependent and reversible. 
Capturing these transformations under reaction conditions is essential, because the active surface is not a static termination of a bulk crystal but a dynamic state that can select pathways and control selectivity, as clearly identified in a recent study of the Pt(100) surface \cite{Simonne2026a}.
Industrial catalysts have long relied on Pt--Rh gauzes, where Rh additions (typically up to \qty{10}{\percent}) improve lifetime and reduce platinum loss \parencite{Kaiser1909, Handforth1934, Heck1982}. 
Under operating conditions, oxygen-rich and high-temperature environments promote substantial surface evolution, including restructuring and roughening, with direct consequences for activity and selectivity \parencite{McCabe1974, Hatscher2008, Simonne2026a}. 
On Pt--Rh(100), reversible relaxation transitions between nitrogen- and oxygen-rich feeds were reported and linked to electronic-structure changes that affect adsorption \parencite{Resta2020a, Hammer1995, Mavrikakis1998, Hammer2000}. 
Rh-containing oxides have further been proposed to act as oxygen reservoirs and to stabilize volatile platinum oxides \parencite{Hatscher2008, Resta2020a}, recent work also showed that Rh strongly impinges on the formation of platinum oxides on Pt(100) that are associated with material loss \parencite{Simonne2026a, Hatscher2008}. 
In particular, the role of Pt-oxide surface phases during ammonia oxidation, and how alloying shifts their stability and kinetics, are still not resolved at the level of detailed structure.
Progress on these questions requires \textit{operando} structural probes that are sensitive to the topmost atomic layers.
A key \textit{operando} study combining surface X-ray diffraction (SXRD) and Bragg coherent diffraction imaging (BCDI) showed that Pt microcrystals reconstruct under oxygen-rich ammonia oxidation, with increased exposure of \{111\} and \{100\} facets, consistent with facet-dependent stabilization of oxidized surfaces predicted by theory \parencite{Simonne2026, Seriani2008}. 
While BCDI provides access to particle shape and internal strain fields, it samples a limited region of reciprocal space and does not directly resolve thin crystalline surface oxides or reconstructions.
In contrast, SXRD offers a unique and powerful route to surface structure under realistic pressures and temperatures: crystal truncation rods (CTRs) quantify relaxations and roughness at the surface, while superstructure features reveal reconstructions and ordered oxide-related phases with surface specificity.
These capabilities make SXRD particularly well suited to link atomic-scale surface structure to catalytic performance in environments where microscopy and ultra-high-vacuum approaches are constrained.
In this work, we present an \textit{operando} investigation of Pt(111) under industrially relevant ammonia oxidation conditions, following our previous study on Pt(100) \cite{Simonne2026a}. 
In-plane reciprocal-space maps track surface reconstructions and the emergence of oxide-related superstructures. 
CTR measurements quantify surface relaxation and roughness, and dedicated superstructure rods probe the out-of-plane structure of newly observed surface phases. 
Complementary near-ambient pressure X-ray photoelectron spectroscopy (NAP-XPS) provides chemical information through O 1s, N 1s, and Pt 4f core-level spectra. 
By systematically varying the ammonia-to-oxygen ratio and simultaneously monitoring products with real-time residual gas analysis (RGA), we relate surface structure and surface chemistry to catalytic selectivity on Pt(111), with the goal of clarifying which dynamic surface states control the reaction mechanism.

\section{Results}

\subsection{Experimental conditions}

\begin{table}[!htb]
\centering
\resizebox{\textwidth}{!}{%
\begin{tabular}{@{}c|c|c|ccc|l@{}}
\toprule
Order & Techniques & Total Pressure (\unit{\milli\bar}) & \ce{Ar} (\unit{\milli\bar}) & \ce{O2} (\unit{\milli\bar}) & \ce{NH3} (\unit{\milli\bar}) & Targeted Information \\
\midrule
1 & SXRD, NAP-XPS & 500 / 1     & 500 / 1   & 0 / 0     & 0 / 0     & Catalyst state without reactants \\
2 & SXRD, NAP-XPS & 500 / 8.8   & 420 / 0   & 80 / 8.8  & 0 / 0     & Growth of surface oxides \\
3 & SXRD, NAP-XPS & 500 / 9.9   & 410 / 0   & 80 / 8.8  & 10 / 1.1  & \multirow{2}{*}{\begin{tabular}[c]{@{}l@{}}Influence of (\ce{NH3}/\ce{O2}) ratio and impact \\ on/of potential surface oxides\end{tabular}} \\
4 & SXRD, NAP-XPS & 500 / 1.65  & 485 / 0   & 5 / 0.55  & 10 / 1.1  & \\
5 & SXRD, NAP-XPS & 500 / 1.1   & 490 / 0   & 0 / 0     & 10 / 1.1  & Ammonia adsorption \\
6 & SXRD, NAP-XPS & 500 / 1     & 500 / 1   & 0 / 0     & 0 / 0     & Returning to pristine state \\
7 & SXRD, NAP-XPS & 500 / 0.55  & 495 / 0   & 5 / 0.55  & 0 / 0     & Growth of surface oxides \\
\bottomrule
\end{tabular}%
}
\caption{
Combined experimental conditions for SXRD and NAP-XPS. All SXRD experiments were conducted at \qty{500}{\milli\bar}; NAP-XPS experiments were performed at lower total pressure to increase the electron counts.
}
\label{tab:Combined_Conditions_SXRD_XPS}
\end{table}

The ammonia oxidation on Pt(111) was first measured \textit{operando} using SXRD, and then with NAP-XPS.
Both experiments were performed at \qty{450}{\degreeCelsius}, a temperature above the catalyst light-off (Figure \ref{fig:TempRamps} - \parencite{Simonne2026, Simonne2026a}), and at different gas atmosphere as summarised in Table \ref{tab:Combined_Conditions_SXRD_XPS}.
Extended details regarding the experimental setups and sample preparation procedure are available in the Methods section.
Argon was used before (1) and after (6) the oxidation cycle to evaluate the reversibility of the catalyst surface.
To ensure that any surface reconstruction/relaxation effects are attributed to the simultaneous presence of ammonia and oxygen in the reactor, the sample was first exposed to a high-oxygen atmosphere (2), monitoring potential surface/bulk oxide growth.
Introducing ammonia (3) facilitated the correlation of such oxides with the catalyst selectivity, while also studying their stability under reaction conditions.
Two different partial pressures of oxygen were employed alongside the same partial pressure of ammonia to investigate the influence of the $p_{O_2}/p_{NH_3}$ ratio on the catalyst structure, surface species, and selectivity (3, 4).
Removing oxygen to keep only ammonia in the reactor (5) enabled the separation of the effect of ammonia's presence from the combined presence of ammonia and oxygen.
Finally, the sample surface was exposed to a reduced pressure of oxygen, to make certain that the observation, when lowering the $p_{O_2}/p_{NH_3}$ ratio, are due to the combined presence of the reactants.
All gases have similar thermal conductivity (Table \ref{tab:InterplanarSpacingsPt111Oxygen}).


\subsection{Surface oxidation at high oxygen partial pressure}

\begin{figure}[!htb]
    \centering
    \includegraphics[width=\textwidth]{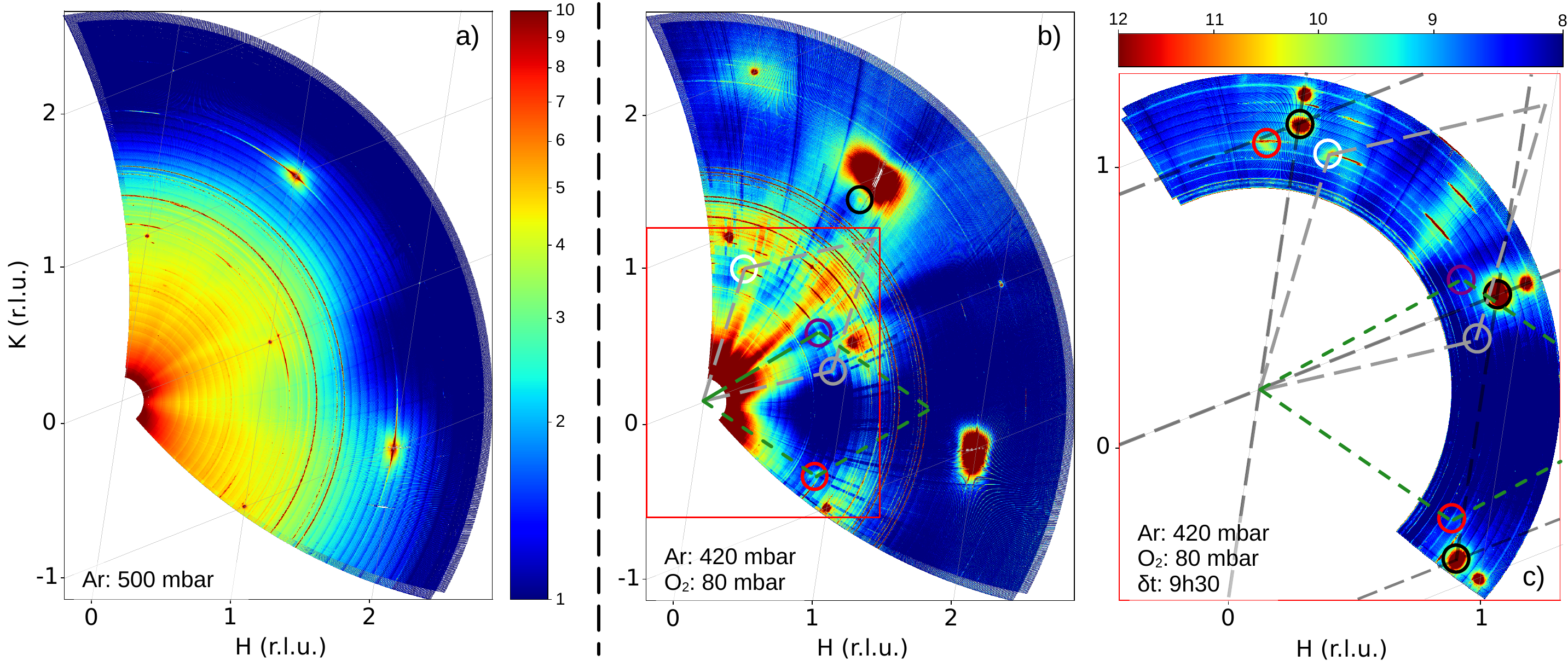}
    \caption{
        Reciprocal space in-plane maps collected under different atmospheres, computed using the hexagonal lattice of Pt(111).
        The measurement of the large reciprocal space in-plane map takes about \qty{3}{\hour}\qty{30}{\minute}, and \qty{1}{\hour}\qty{15}{\minute} for the small map.
        Both measurements under oxygen share the same colormap.
    }
    \label{fig:MapsPt111A}
\end{figure}

The first reciprocal space in-plane map is under inert atmosphere  (Figure \ref{fig:MapsPt111A}a), collected after sputtering and annealing under high vacuum (see Methods).
All reciprocal space maps in this study are measured at $L = 0.15$ and at \qty{450}{\degreeCelsius}.
The (110) and (2$\bar{1}$0) Bragg peaks are visible, together with the intersection of the [0, 1, L], [0, 2, L], [1, $\bar{1}$, L], [1, 0, L], and [2, 0, L] CTRs with the [H, K, 0] plane.
The sample alignment was corrected after the first map, as it seems that thermal stability was not yet reached from the decreasing intensity gradient (the measurement starting from the reciprocal space centre).
New peaks are first observed during surface oxidation at $p_{O2} = \qty{80}{\milli\bar}$ (Figure \ref{fig:MapsPt111A}b).
All corresponding interplanar spacings are given in Table \ref{tab:InterplanarSpacingsPt111Oxygen}.
The angle between the peaks circled in grey and white, and the peaks circled in red and purple is equal to \ang{60}, proving the existence of two hexagonal surface structures, each one rotated by \ang{\pm 8.8} with respect to the Pt(111) surface unit cell and with a larger in-plane lattice parameter (drawn in green and grey in Figure \ref{fig:MapsPt111A}b-c).
However, no apparent second order peak can be linked to those rotated hexagonal structures, the related intensity is possibly too low at high scattering vector to be detected.
The closest rotated and commensurate superstructure is Pt(111)-($6\times6$)-R\ang{\pm 8.8}.
Interestingly, summing two vectors each going from the centre of the reciprocal space to either the white or purple circled peaks yields the position of the black circled peak in Figure \ref{fig:MapsPt111A}b.
The angle between both vectors is then equal to \ang{42.6}, equivalent to \ang{137.4} in real space.
No other peak linked to this unit cell can be identified in this in-plane map.
A second map was measured in a smaller region of the reciprocal space \qty{9}{\hour}\qty{30}{\minute} after the introduction of oxygen in the cell.
Three additional peaks (circled in black in Figure \ref{fig:MapsPt111A}c) describe an epitaxied hexagonal lattice with $|\vec{a_{hex}}| = \qty{3.12}{\angstrom}$, the closest commensurate superstructure is Pt(111)-($8\times8$).
The grey and purple circled peaks have vanished from the reciprocal map, later measurements show that this is due to alignment issues during the oxidation reaction.

\begin{figure}[!htb]
    \centering
    \includegraphics[width=\textwidth]{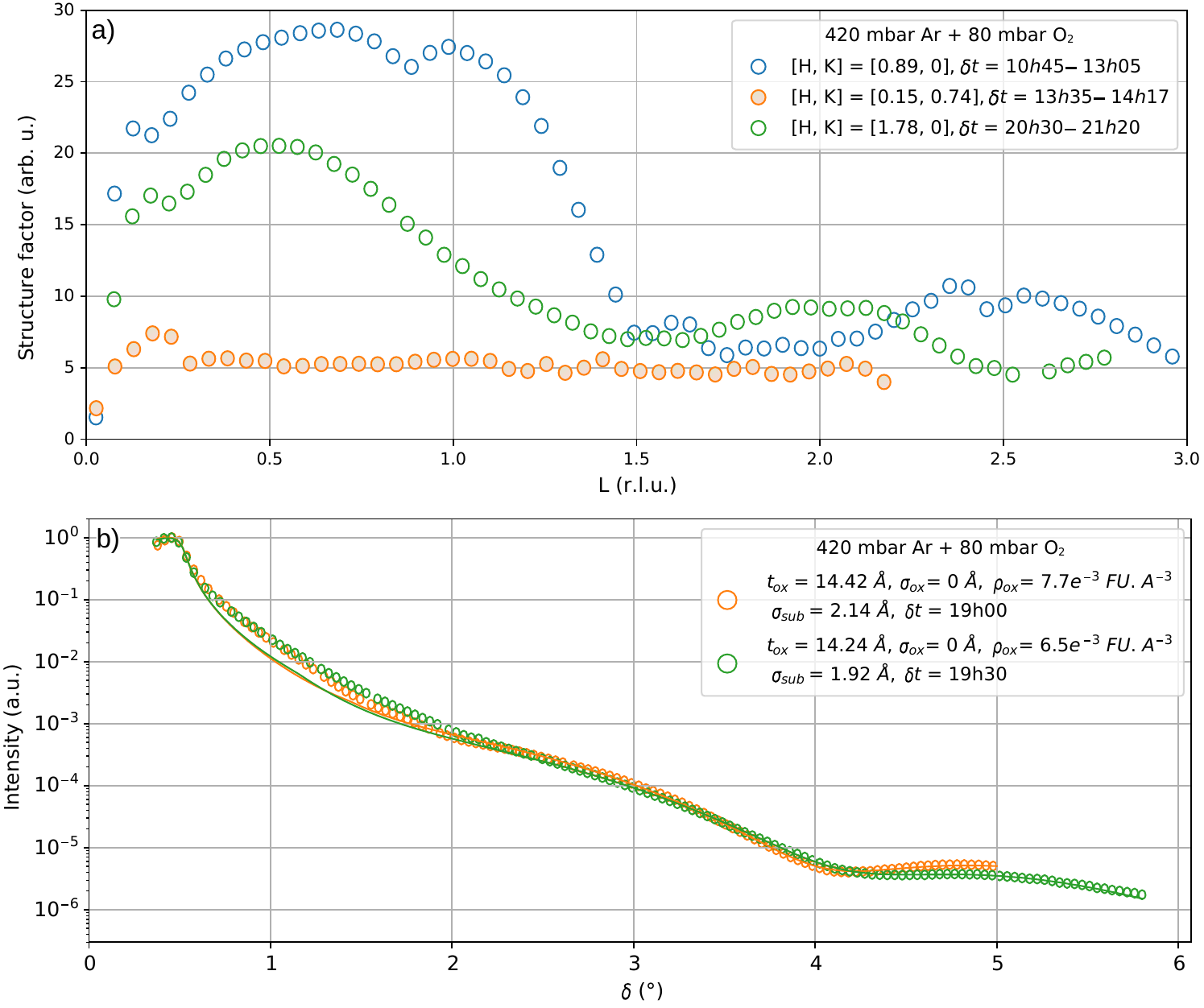}
    \caption{
        Super-structure rods (SSRs) measurements for three different [H, K] positions at $p_{Ar} = \qty{420}{\milli\bar}$ and $p_{O_2} = \qty{80}{\milli\bar}$ (a)
        $\delta t$ designates the elapsed time since the introduction of oxygen, until the end of the measurement. 
        The empty circles correspond to the Pt(111)-($8\times8$) structure.
        X-ray reflectivity curves at $p_{Ar} = \qty{420}{\milli\bar}$ and $p_{O_2} = \qty{80}{\milli\bar}$ for different exposure times (b)
    	Curves fitted using \text{GenX} are shown as lines.
        $\delta t$ designates the time elapsed since the introduction of oxygen.
        $\rho_{ox}$, $t_{ox}$, and $\sigma_{ox}$ are the oxide density, thickness, and root mean square roughness.
        $\sigma_{sub}$ is the substrate root mean square roughness.
    }
    \label{fig:LScansAndReflecto80}
\end{figure}

Super-structure rods (SSRs) were collected on the peaks belonging to the Pt(111)-($8\times 8$) structure at [H, K] = [0.89, 0], and to the rotated hexagonal structures at [H, K] = [0.15, 0.74], to probe their out-of-plane order.
A second SSR was measured at [H, K] = [1.78, 0] to probe for a potential second-order peak associated to the Pt(111)-($8\times8$) structure, approximately \qty{10}{\hour} after the measurement at [H, K] = [0.89, 0].
The integrated intensity is presented in Figure \ref{fig:LScansAndReflecto80}a.
The intensity distribution for both Pt(111)-($8\times8$) measurements suggests the presence of a multi-layer structure, in this case a surface oxide, while the small peak at $L\approx 0.2$ followed by a constant intensity for the SSR measured at [H, K] = [-0.15, -0.74] is characteristic of a monolayer / chemisorbed specie \parencite{Robinson1991}.
Reflectivity curves measured in a specular geometry (Figure \ref{fig:LScansAndReflecto80}b) were fitted using \textit{GenX} \parencite{Bjorck2007, Glavic2022} (see Methods).
The position of the oscillations can be linked to the thickness of the oxide layer ($t_{ox}$), \qty{14.42}{\angstrom} and \qty{14.24}{\angstrom} thick in the first and second reflectivity curves, \textit{i.e.} stable between both measurements.

\subsection{Absence of surface reconstructions and roughness decrease during ammonia oxidation}

\begin{figure}[!htb]
    \centering
    \includegraphics[width=\textwidth]{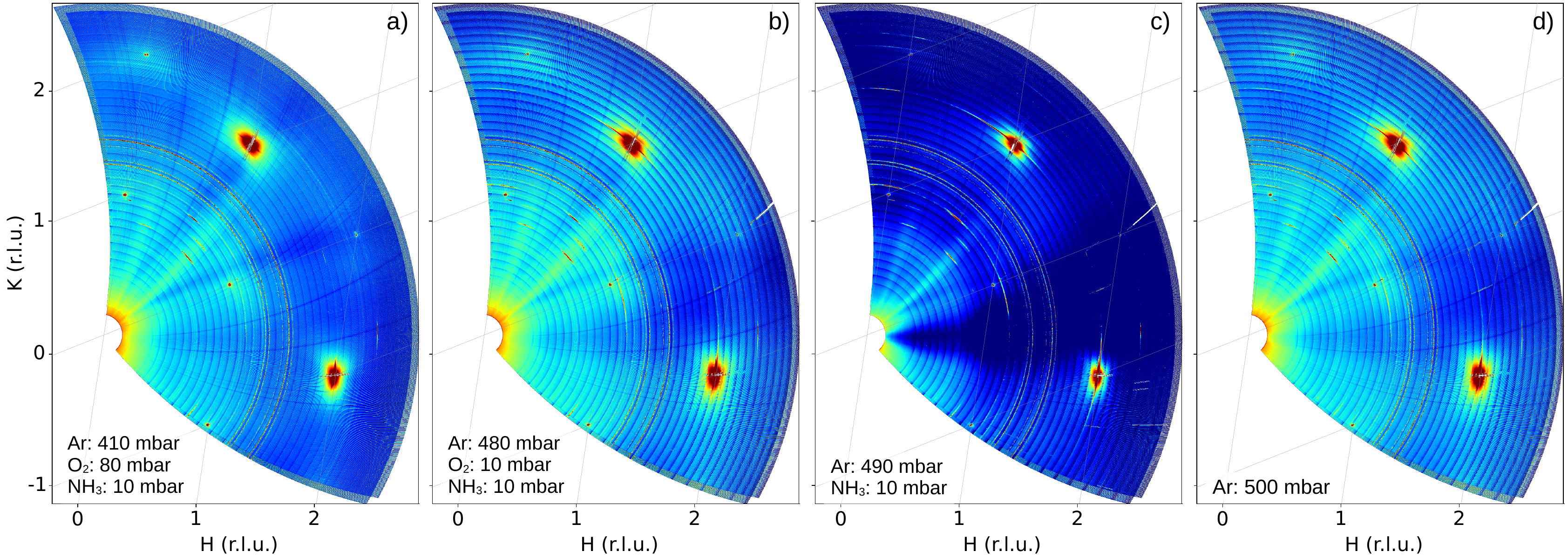}
    \caption{
        Reciprocal space in-plane maps collected under different atmospheres, computed using the hexagonal lattice of Pt(111).
    }
    \label{fig:MapsPt111B}
\end{figure}

\begin{figure}[!htb]
    \centering
    \includegraphics[width=\textwidth]{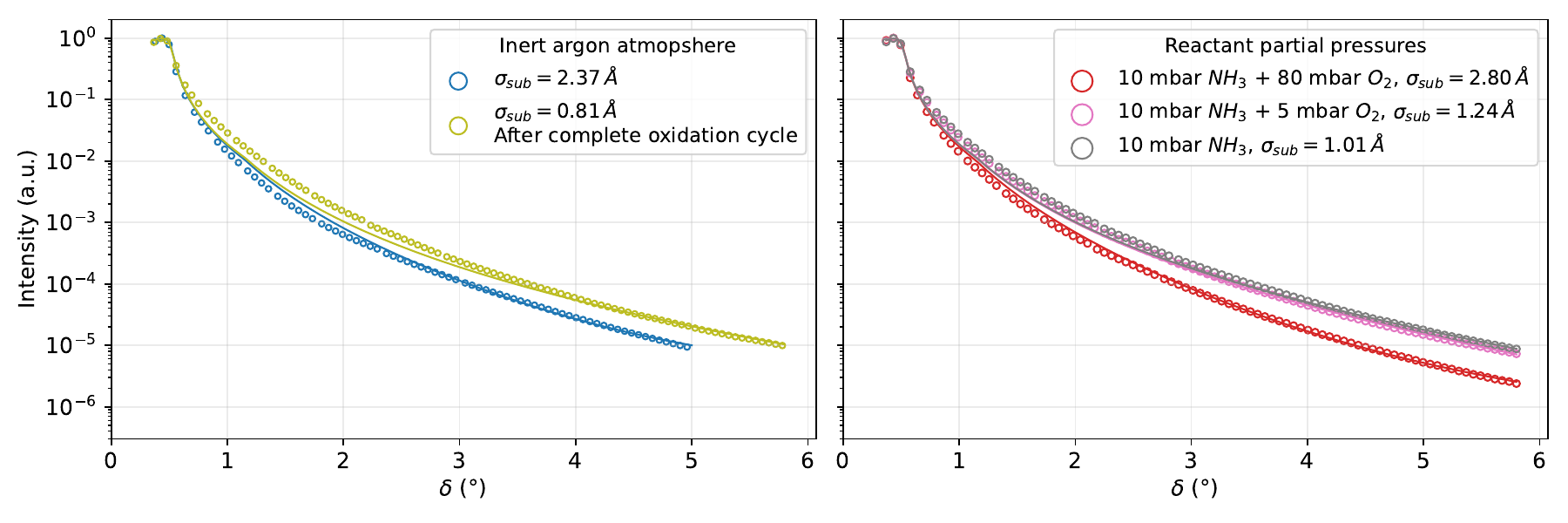}
    \caption{
    	X-ray reflectivity curves under different atmospheres during ammonia oxidation.
        Curves fitted using \text{GenX} are shown as lines.
        $\sigma_{sub}$ is the substrate root mean square roughness.
        }
    \label{fig:ReflectoCycle}
\end{figure}

Ammonia was introduced in the reactor after surface oxidation to investigate the stability of the platinum oxide structures, as well as the potential presence of additional surface structures or surface reconstructions during the reaction (Figure \ref{fig:MapsPt111B}a-b).
None of the previously observed features were measured at $p_{O_2} / p_{NH_3} = 8$ or $p_{O_2} / p_{NH_3} = 0.5$, while no additional peaks emerged, the reaction removes the oxide layer grown during initial surface oxidation.
A gradual increase of \ce{NO} is observed in the first hours following the introduction of ammonia in the reactor (Figure \ref{fig:RGA450Pt111}), the presence of the surface oxide at the beginning of the condition possible influencing the product selectivity.
All reciprocal space in-plane maps during the reaction cycle and back to argon showed a bulk-terminated surface (Figure \ref{fig:MapsPt111B}a-b-c-d) that resembles the first map also measured under argon before surface oxidation (Figure \ref{fig:MapsPt111A}a), proving that the oxidation cycle reproduces the same surface termination.
Complementary X-ray reflectivity curves are presented in Figure \ref{fig:ReflectoCycle}.
In accordance with the results of the in-plane reciprocal space maps in Figure \ref{fig:MapsPt111B}, no oxide layer was used in the fitting model.
The surface roughness increased to \qty{2.80}{\angstrom} following the introduction of ammonia (Figure \ref{fig:ReflectoCycle}b), after reaching \qty{2.14}{\angstrom} when $p_{O_2} = \qty{80}{\milli\bar}$ (Figure \ref{fig:LScansAndReflecto80}).
Since the X-ray reflectivity curves are measured at the start of each condition, the high roughness when $p_{O_2} / p_{NH_3} = 8$ can be linked to the removal of the surface oxides.
Lowering the amount of oxygen decreases the roughness (Figure \ref{fig:ReflectoCycle}b).
The largest decrease is observed when switching the $p_{O_2}/p_{NH_3}$ ratio from 8 to 0.5, \textit{i.e.} during reacting conditions.

\subsection{Surface oxidation at low oxygen partial pressure}

\begin{figure}[!htb]
    \centering
    \includegraphics[width=\textwidth]{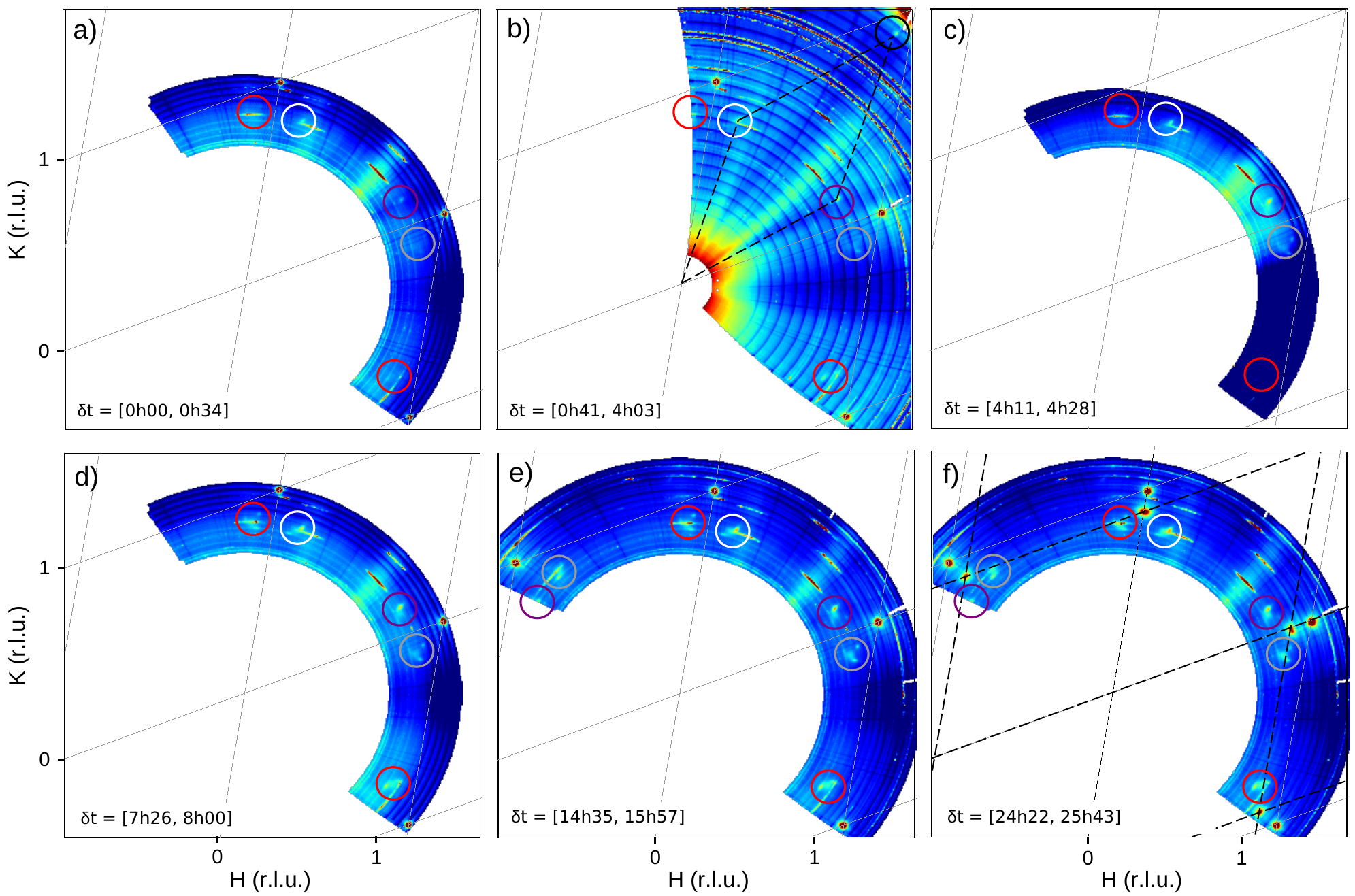}
    \caption{
        Large reciprocal space in-plane maps collected at $p_{Ar} = \qty{495}{\milli\bar}$ and $p_{O_2} = \qty{5}{\milli\bar}$ for different exposure times.
    }
    \label{fig:LargeMapsPt111LowOxygen}
\end{figure}

At the end of the ammonia oxidation cycle, the sample was cleaned with two sputtering and annealing cycles and put under \qty{5}{\milli\bar} of oxygen to monitor the growth of surface oxides at lower oxygen pressure.
Large and small reciprocal space in-plane maps were measured to have the best compromise between a higher temporal resolution of the oxide growth, and the possibility to detect other peaks from corresponding surface unit cells.
The larger reciprocal space in-plane maps are shown in Figure \ref{fig:LargeMapsPt111LowOxygen}.
The first map starts directly after the stabilisation of the oxygen pressure at \qty{5}{\milli\bar} in the reactor cell.
Low intensity signals are instantly detected (Figure \ref{fig:LargeMapsPt111LowOxygen}a), corresponding to those previously measured shortly after setting a pressure of \qty{80}{\milli\bar} in the cell (Figure \ref{fig:MapsPt111A}c), proving that ammonia effectively impinges on their formation at both reactant pressure ratios.
Circles of the same colour are used to indicate peaks at similar positions, the related interplanar spacing is also summarised in Table \ref{tab:InterplanarSpacingsPt111Oxygen}.
Some peaks split in a similar pattern, exhibiting three distinct signals around a more diffuse region (clearly visible for the red circled peak in Figure \ref{fig:LargeMapsPt111LowOxygen}d).
This splitting can be interpreted as the signature of a Pt(111) surface covered by different domains that exhibit a similar hexagonal in-plane symmetry.
These unit cells are on average rotated by \ang{\pm 8.8} with respect to the Pt(111) surface unit cell, with a higher magnitude of the in-plane lattice parameter, following the same Pt(111)-($6\times6$)-\ang{\pm8.8} unit cells observed under an oxygen pressure of \qty{80}{\milli\bar} (Figure \ref{fig:MapsPt111A}c).
No corresponding second order peak is observed.
The grey circled peak becomes fully visible after \qty{4}{\hour}, which confirms that the absence of peak in Figure \ref{fig:MapsPt111A}c come from alignments problems.
A peak can be observed at [H, K] = [0.89, 0.89] in the large reciprocal space in-plane map collected between \qty{41}{\minute} and \qty{4}{\hour}\qty{3}{\minute} (circled in black in Figure \ref{fig:LargeMapsPt111LowOxygen}b).
Its position is the same as in the first large reciprocal space in-plane map measured at $p_{O_2} = \qty{80}{\milli\bar}$ (Figure \ref{fig:MapsPt111A}c), and corresponds to the Pt(111)-($8\times8$) commensurate superstructure.
However, no peaks at [0.89, 0, 0] and [0, 0.89, 0] can yet be detected, before (Figure \ref{fig:MapsPt111A}b), and after (Figure \ref{fig:MapsPt111A}c) measuring this signal.
Such peaks are only detected in the large maps after \qty{\approx 24}{\hour} of measurements (described by the hexagonal lattice drawn with black dashed lines in Figure \ref{fig:MapsPt111A}f).
As mentioned before, it is possible to draw a surface unit cell in Figure \ref{fig:MapsPt111A}b that includes the peak at [0.89, 0.89] together with the white and purple circled peaks.
The area sampled during the measurement was extended in the last two maps to observe more signals around the ($\bar{1}10$) region (Figure \ref{fig:LargeMapsPt111LowOxygen}e-f).
For each map, the purple and red circled peaks around the Pt(111)-($8\times8$) peaks are separated by \ang{60}, likewise for the grey and white circled peaks.
In the last two maps, four doublet of peaks are visible, all split in a similar pattern.
During this set of measurement, the grey circled peak did not disappear in contrast with the measurements carried out at $p_{O_2} = \qty{80}{\milli\bar}$, which furthermore supports the existence of rotated hexagonal structures that coexist with the Pt(111)-($8\times8$) structure.

\begin{figure}[!htb]
    \centering
    \includegraphics[width=\textwidth]{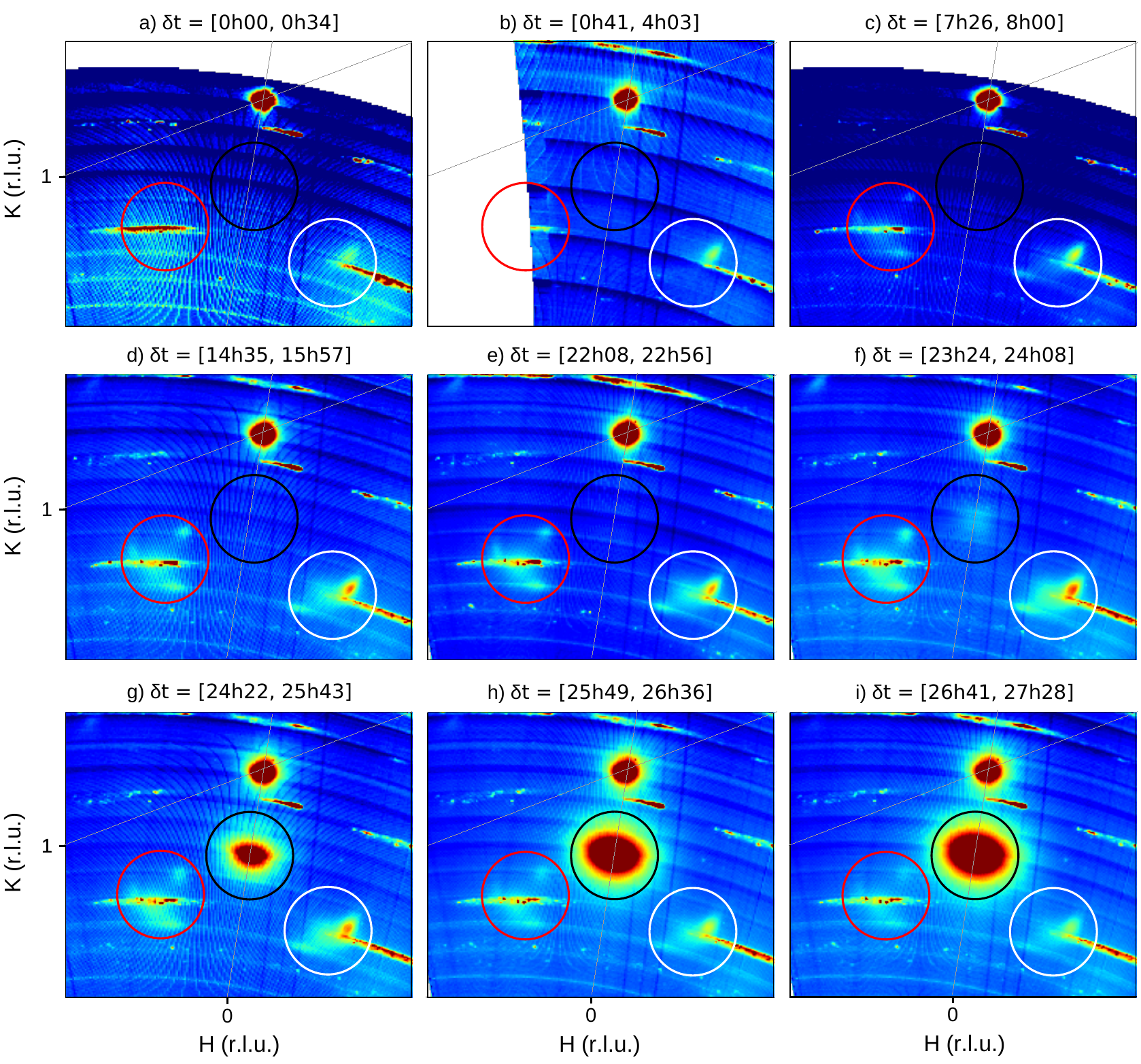}
    \caption{
        Small reciprocal space in-plane maps collected at $p_{Ar} = \qty{495}{\milli\bar}$ and $p_{O_2} = \qty{5}{\milli\bar}$ for different exposure times.
    }
    \label{fig:SmallMapsPt111LowOxygen}
\end{figure}

\begin{figure}[!htb]
    \centering
    \includegraphics[width=\textwidth]{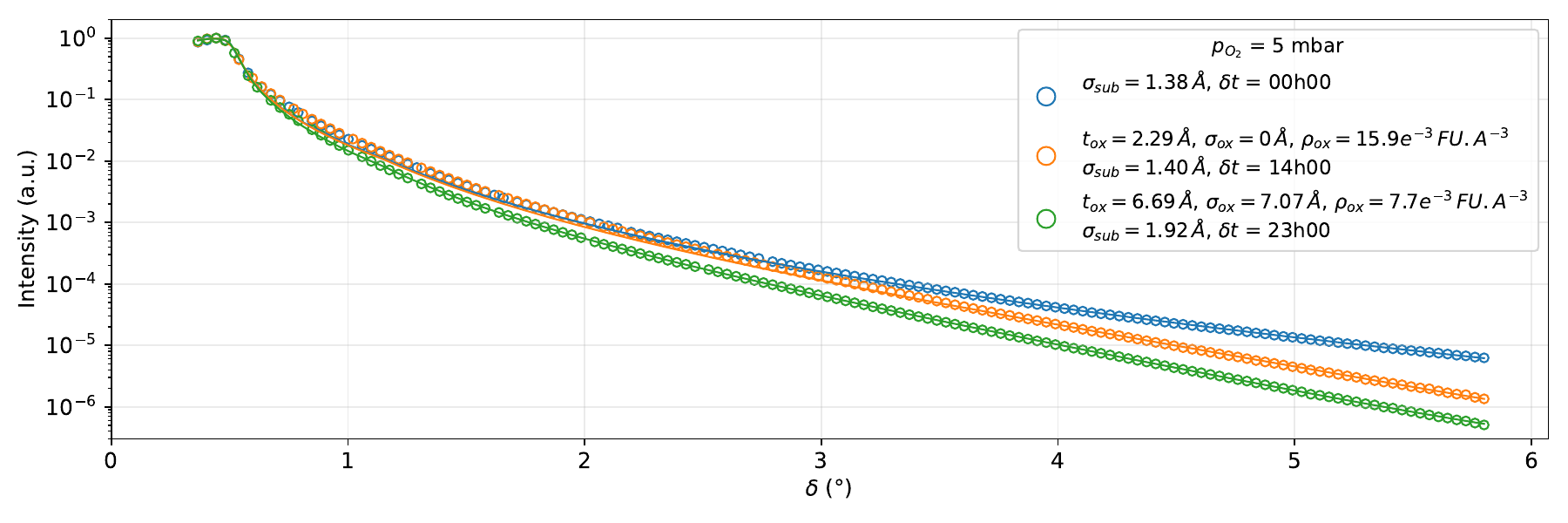}
    \caption{
    	X-ray reflectivity curves measured in a specular geometry (full lines)at $p_{Ar} = \qty{495}{\milli\bar}$ and $p_{O_2} = \qty{5}{\milli\bar}$ for different exposure times.
        Curves fitted using \text{GenX} are shown as lines.
        $\delta t$ designates the elapsed time since the introduction of oxygen.
        $\rho_{ox}$, $t_{ox}$, and $\sigma_{ox}$ are the oxide density, thickness, and root mean square roughness.
        $\sigma_{sub}$ is the substrate root mean square roughness.
    }
    \label{fig:Reflecto5}
\end{figure}

Smaller maps taken with a shorter time interval are presented in Figure \ref{fig:SmallMapsPt111LowOxygen}, a Pt(111)-($8\times8$) peak is detected after \qty{\approx 23}{\hour} of measurements at [0.89, 0.89] (circled in black).
The same peak was measured after only \qty{13}{hour} when $p_{O_2} = \qty{80}{\milli\bar}$ (Figure \ref{fig:MapsPt111A}d), highlighting the importance of the oxygen chemical potential in the surface oxidation.
Finally, X-ray reflectivity curves were measured before, \qty{14}{\hour}, and \qty{23}{\hour} after the introduction of oxygen (Figure \ref{fig:Reflecto5}).
The last reflectivity curve was measured just before the detection of the Pt(111)-($8\times8$) structure with SXRD (Figure \ref{fig:SmallMapsPt111LowOxygen}f).
Following the information gathered by SXRD, a homogeneous layer on top of Pt(111) was introduced in the fitting model for the measurements at \qty{14}{\hour} and \qty{23}{\hour}, to represent the rotated structures visible in Figure \ref{fig:SmallMapsPt111LowOxygen}.
The oxide layer was necessary for the \qty{23}{\hour} curve to achieve a good fit, but only slightly improved the result for the \qty{14}{\hour} curve.
The layer thickness and density are difficult to estimate since no oscillations are visible, possibly too far in $\delta$.
However, this confirms that the oscillations measured at $p_{O_2} = \qty{80}{\milli\bar}$ in Figure \ref{fig:LScansAndReflecto80} are linked to the Pt(111)-($8\times8$) structure and not to the rotated hexagonal structures, and that a longer exposure time is needed to reach the same thickness.
This also further confirms that the red and white peaks are linked to monolayers (also confirmed by the lack of peak in the related SSRs in Figure \ref{fig:LScans05}).
A transition in terms of oxide roughness and density is observed between the orange and green curves.
Interestingly, the final density is equal to that retrieved at $p_{O_2} = \qty{80}{\milli\bar}$ (Figure \ref{fig:LScansAndReflecto80}), but without the same layer thickness, and roughness.
This effect is clearly visible in the reflectivity curve with a decrease of intensity starting at low $\delta$ angles.
The substrate roughness increases with the elapsed time since the introduction of oxygen, coherent with the creation of a surface oxide.

\subsection{Surface roughness and surface relaxation effects}

\begin{figure}[!htb]
    \centering
    \includegraphics[page=1, width=\textwidth]{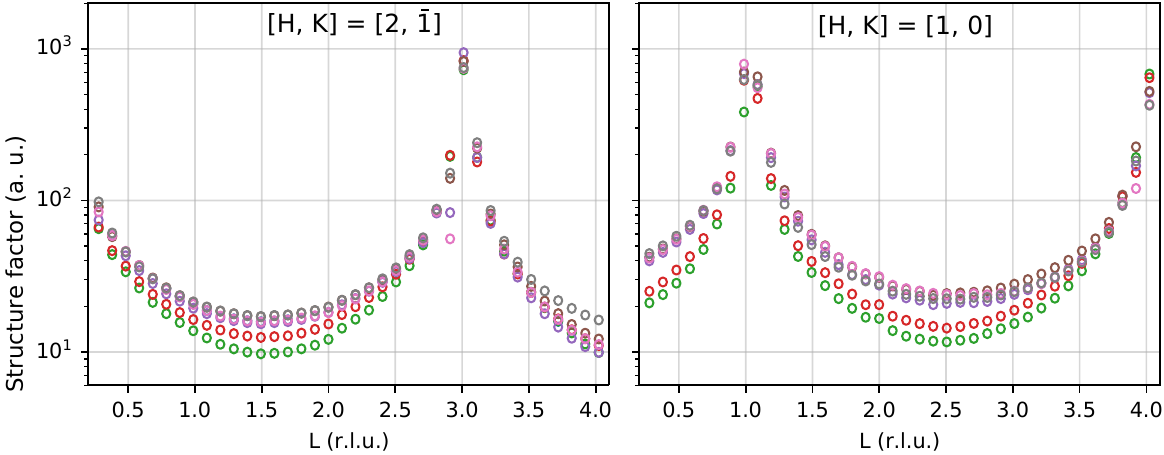}
    \vspace{0.1cm}
    \includegraphics[page=2, width=\textwidth]{Figures/SXRD/ctr_together.pdf}
    \caption{
        Evolution of crystal truncation rods under different atmospheres.
        Only the ($2\bar{1}L$) CTR was measured under \qty{500}{\milli\bar} of argon before the oxidation cycle.
    }
    \label{fig:CTRPt111}
\end{figure}

The ($2\bar{1}L$), (10L) and (11L) crystal truncation rods intensities are presented in Figure \ref{fig:CTRPt111}.
The evolution of the surface roughness when changing condition is visible from the variation of intensity minimum near $L=1.5$ (or $L=2.5$ for the (10L) CTR).
Overall, the intensity distribution is symmetric around those minima, the surface layers lattice parameters are thus close to their bulk values \parencite{Robinson1986}.

\subsection{Surface species}

The mass spectrometer signals from the NAP-XPS experiment are available in Figure \ref{fig:XPS111RGA}. 
Because the XPS mass‑spectrometer measurements are taken from gases pumped trough the inlet of the XPS analyser, which is closer to the sample surface, it is more representative of the catalyst surface activity. 
The mass spectrometry data from the SXRD experiment instead reflect the average composition of the gas in the reactor.
The production of \ce{N2} and \ce{H2} under pure ammonia measured during the XPS experiment are therefore more visible on average than with the SXRD setup (Figure \ref{fig:RGA450Pt111}).
The high $p_{O_2}/p_{NH_3}$ ratio favours the production of \ce{NO} as expected \parencite{Hatscher2008}, accompanied by water, \ce{N2O} and \ce{N2} in low amounts.
Despite the relative high amount of oxygen in the reactor, ammonia is still detected.
The complete oxidation reaction may be limited by the availability of active sites, the transportation of reactants towards the catalyst surface or the relatively low temperature compared to industrial conditions (\qty{>700}{\degreeCelsius}) which can for example impact the active sites turnover frequency.
Reducing the $p_{O_2}/p_{NH_3}$ ratio to \num{0.5} by lowering the oxygen pressure shifts the reaction selectivity entirely towards \ce{N_2}, water is still the main product.
Oxygen is undetected near the catalyst surface, fully consumed for the production of \ce{N_2} and \ce{H_2O} \textit{via} the oxidation reaction.
Ammonia can be thus considered to be in excess, a new production of \ce{H_2} is measured, which supports that the simultaneous dissociation of \ce{NH_3} is favoured is this condition.
When removing oxygen, only the dissociation of ammonia is observed, put into evidence by the presence of hydrogen and absence of water.

\subsubsection*{O 1s  and N 1s Spectra: oxygen speciation and nitrogen adsorbates}

\begin{figure}[!htb]
    \centering
    \includegraphics[width=\textwidth]{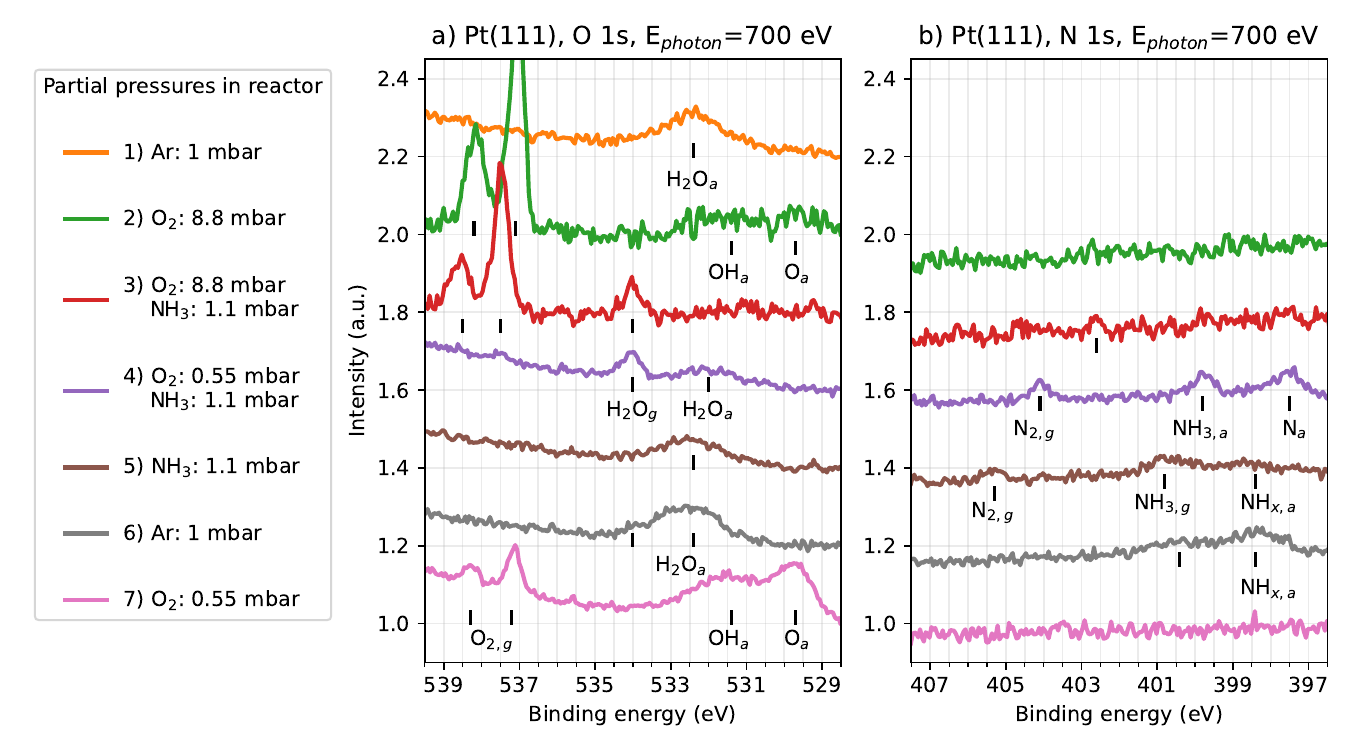}
    \caption{
        XPS spectra collected at the O 1s (a) and N1 s (b) levels under different atmospheres at \qty{450}{\degreeCelsius} with an incoming photon energy of \qty{700}{\eV}.
        The spectra are normalised and shifted in intensity to highlight the presence of different peaks.
    }
    \label{fig:O1sN1sPt111}
\end{figure}

The evolution of the N 1s and O 1s XPS spectra for different atmospheres is displayed in Figure \ref{fig:O1sN1sPt111}.
Binding energy are given with reference to the Fermi level, and re-normalised at the pre-edge intensity (low binding energy side.
All the reported peaks and corresponding species in the discussion below are summarised in Table \ref{tab:XPSPt111}.
Subscript \textit{a} and \textit{g} stand for adsorbed and gas phase respectively.

\begin{table}[!htb]
\centering
\resizebox{\textwidth}{!}{%
    \begin{tabular}{@{}ll|lllllll@{}}
    \toprule
    \multirow{3}{*}{Partial pressures (\unit{\milli\bar})} & \ce{Ar}   & 1 & 0   & 0   & 0    & 0   & 1 & 0    \\
                                              & \ce{NH_3} & 0 & 0   & 1.1 & 1.1  & 1.1 & 0 & 0    \\
                                              & \ce{O_2}  & 0 & 8.8 & 8.8 & 0.55 & 0   & 0 & 0.55 \\
    \midrule
    Gas signals (decreasing & & Ar & \ce{O_2} & \ce{O_2}, \ce{H_2O}, \ce{NO}   & \ce{H_2O}, \ce{NH_3} & \ce{H_2}, \ce{NH_3} & Ar & \ce{O_2} \\
    pressure order)          & &    &          & \ce{NH_3}, \ce{N_2}, \ce{N_2O} & \ce{N_2}, \ce{H_2}   & \ce{N_2}            &    &          \\
    \midrule
    \multicolumn{2}{l|}{N 1s: peak positions}
        & No data          & No peak          & \qty{402.6}{\eV} & \qty{404.1}{\eV} & \qty{405.3}{\eV} & \qty{400.4}{\eV} & No peak          \\
     &  &                  &                  &                  & \qty{399.8}{\eV} & \qty{400.8}{\eV} & \qty{398.4}{\eV} &                  \\
     &  &                  &                  &                  & \qty{397.5}{\eV} & \qty{398.4}{\eV} &                  &                  \\
    \multicolumn{2}{l|}{Attributed surface species}
        &                  &                  & Not indexed      & \ce{N_{2,g}}     & \ce{N_{2,g}}     & \ce{NH_{3,a}}    &                  \\
     &  &                  &                  &                  & \ce{NH_{3,a}}    & \ce{NH_{3,g}}    & \ce{NH_{x,a}}    &                  \\
     &  &                  &                  &                  & \ce{N_a}         & \ce{NH_{x,a}}    &                  &                  \\
    \midrule
    \multicolumn{2}{l|}{O 1s: peak positions}
        & \qty{532.4}{\eV} & \qty{538.2}{\eV} & \qty{538.5}{\eV} & \qty{534.0}{\eV} & \qty{532.4}{\eV} & \qty{534.0}{\eV} & \qty{538.3}{\eV} \\
     &  &                  & \qty{537.1}{\eV} & \qty{537.5}{\eV} & \qty{532.0}{\eV} &                  & \qty{532.4}{\eV} & \qty{537.2}{\eV} \\
     &  &                  & \qty{531.4}{\eV} & \qty{534.0}{\eV} &                  &                  &                  & \qty{531.4}{\eV} \\
     &  &                  & \qty{529.7}{\eV} &                  &                  &                  &                  & \qty{529.7}{\eV} \\
    \multicolumn{2}{l|}{Attributed surface species}
        & \ce{H_2O_a}      & \ce{O_{2,g}}     & \ce{O_{2,g}}     & \ce{H_2O_g}      & \ce{H_2O_a}      & \ce{H_2O_g}      & \ce{O_{2,g}}     \\
     &  &                  & \ce{O_{2,g}}     & \ce{O_{2,g}}     & \ce{H_2O_a}      &                  & \ce{H_2O_a}      & \ce{O_{2,g}}     \\
     &  &                  & \ce{OH_a}        & \ce{H_2O_g}      &                  &                  &                  & \ce{OH_a}        \\
     &  &                  & \ce{O_a}         &                  &                  &                  &                  & \ce{O_a}         \\
    \bottomrule
    \end{tabular}%
    }
    \caption{Indexing of peaks measured during ammonia oxidation of the Pt(111) surface.}
\label{tab:XPSPt111}
\end{table}

\begin{description}
    \item[Argon:] A peak is visible at \qty{532.4}{\eV} in the O 1s edge, attributed to adsorbed water \parencite{Fisher1980, Kiskinova1985}, possibly from contaminants in the gas line.
    
    \item[High oxygen:] The oxidation cycle is started by introducing \qty{8.8}{\milli\bar} of oxygen in the reactor.
    The presence of gas phase oxygen (\ce{O_{2,g}}) can be confirmed in the O 1s spectra by a characteristic peak doublet around \qty{538}{\eV}.
    The positions are shifted by \qty{2.2}{\eV} in binding energy with respect to literature (\qty{539.3}{\eV}, \qty{540.4}{\eV}, Avval et al. \cite*{Avval2022}).
    This is explained by a change in the work function of the sample-analyser setup, impacting gas species exclusively since not in electrical contact with the system \parencite{Starr2021}.
    The broad signal extending from \qtyrange{528.5}{533}{\eV} is hiding various peaks from oxygen species.
    Fisher et al \parencite*{Fisher1980} report the signal for adsorbed oxygen (\ce{O_{a}}) and adsorbed hydroxyl groups (\ce{OH_{a}}) at respectively \qty{529.8}{\eV} and \qty{531}{\eV} when exposing the Pt(111) surface to water.
    Peuckert et al. \parencite*{Peuckert1984} have studied various oxidised Pt surfaces and indexed a peak at \qty{530.2}{\eV} for \ce{O_{a}} on Pt(111), and at \qty{531.5}{\eV} for \ce{OH_{a}} on poly-crystalline Pt.
    Derry et al. \parencite*{Derry1984} report \qty{530.8}{\eV} for \ce{O_{a}} on Pt(111) during its exposition to oxygen, while Zhu et al. \parencite*{Zhu2003} report \ce{O_{a}} at \qty{529.9}{\eV} when probing the dissociation of \ce{NO} on the Pt(111) surface.
    Fantauzzi et al. \parencite*{Fantauzzi2017} report oxygen surface species at \qty{529.7}{\eV} during the oxidation of Pt(111) at \qty{225}{\degreeCelsius}, similarly to Miller et al. \parencite*{Miller2014}.
    During a recent study of the oxidation of ammonia at different pressures and \ce{O_2}/\ce{NH_3} ratio on Pt(111), ($2\times2$) chemisorbed oxygen and hydroxyl groups were reported respectively at \qty{529.7}{\eV} and \qty{531.4}{\eV} in \qty{1}{\milli\bar} of oxygen at \qty{325}{\degreeCelsius} \parencite{Ivashenko2021}.
    However, such surface reconstructions were not observed in the in-plane reciprocal space maps (Figure \ref{fig:MapsPt111A}b-c).
    It is probable that hydroxyl groups as well as atomic oxygen are adsorbed on the surface around \qty{531.4}{\eV} and \qty{529.7}{\eV}, respectively.
    
    \item[High oxygen reacting conditions:] Introducing \qty{1.1}{\milli\bar} of ammonia leads to the $p_{O_2}/p_{NH_3} = 8$ condition.
    Only a weak signal at \qty{402.6}{\eV} can be detected in the N 1s spectrum.
    Gas phase nitric oxide (\ce{NO_{g}}), ammonia (\ce{NH_{3,g}}), nitrogen (\ce{N_{2,g}}), and nitrous oxide (\ce{NO_{2,g}}) are visible in the RGA signals (Figure \ref{fig:XPS111RGA}).
    The main product, \ce{NO_{g}}, is expected between \qty{404.5}{\eV} and \qty{406.7}{\eV} in the N 1s level, observed during reacting conditions with an equal amount of oxygen and ammonia, a total pressure of \qty{1}{\milli\bar}, and temperature of \qty{325}{\degreeCelsius} \parencite{Ivashenko2021}.
    Since not observed in the O 1s level, the authors have proposed that the related signal is hidden by the important gas phase oxygen features.
    The same study reports \ce{N_{2,g}} between \qty{403.9}{\eV} and \qty{404.8}{\eV}, and \ce{NH_{3,g}} at \qty{400.4}{\eV}.
    \ce{NH_{3,g}} was also measured shifted to \qty{400.7}{\eV} under \qty{1}{\milli\bar} of ammonia, without oxygen present in the cell.
    Adsorbed nitrogen oxide (\ce{NO_a}) is absent, expected to yield peaks between \qty{530}{\eV} and \qty{532}{\eV} in the O 1s level, and between \qty{400.4}{\eV} and \qty{401.3}{\eV} in the N 1s level \parencite{Kiskinova1984, Zhu2003, Gunther2008}.
    It is worth noticing that this region is also populated by water and \ce{OH_a} signals.
    Since \ce{NO} is one of the main products at this condition, the absence of \ce{NO_a} can be explained by short desorbing time from the Pt(111) surface \parencite{Ivashenko2021}.
    The peak at \qty{402.6}{\eV} in this study could not be indexed, situated far away in binding energy from any of the reported gas phase peaks, and of adsorbed nitrogen and \ce{NH_x} type features.
    Gas phase water (\ce{H_2O_{g}}) is visible in the O 1s spectra by a peak at \qty{534}{\eV}, as reported during ammonia oxidation by Weststrate et al. \parencite*{Weststrate2006}, generated by the oxidation reaction.
    The energy difference between the low energy peak of \ce{O_{2,g}} and \ce{H_2O_{g}}, \qty{3.5}{\eV}, is close to the difference reported in literature for pure gas phases, equal to \qty{3.3}{\eV} \parencite{Linford2019, Avval2022}.

    \item[Low oxygen reacting conditions:] By lowering the oxygen pressure to \qty{0.55}{\milli\bar} so that $p_{O_2}/p_{NH_3} = 0.5$, nitrogen production is favoured (Figure \ref{fig:XPS111RGA}), three peaks can be detected in the N 1s level.
    \qty{397.5}{\eV} is characteristic of adsorbed atomic nitrogen (\ce{N_{a}}) on Pt(111) \parencite{vandenBroek1999, Zhu2003}.
    The energy difference between \ce{N_{a}}, \ce{NH_{a}}, and \ce{NH_{2,a}} is reported to be approximately \qty{0.95}{\eV}, \qty{1.9}{\eV} in total on Pt(111) \parencite{Ivashenko2021}, which is too few to link the peak at \qty{399.8}{\eV} to \ce{NH_{a}} or \ce{NH_{2,a}} with respect to the \ce{N_{a}} peak.
    Since there is a peak at \qty{400.8}{\eV} in the following spectrum that can be linked to gas phase ammonia from similar reported binding energies, the peak at \qty{399.8}{\eV} is attributed to adsorbed ammonia (\ce{NH_{3,a}}).
    The peak at \qty{404.1}{\eV} is attributed to \ce{N_{2,g}}, which is the only measured nitrogen product, and also in accordance with reported binding energies.
    No more oxygen is detected in the reactor (Figure \ref{fig:XPS111RGA}).
    Moreover, no \ce{OH_{a}} and \ce{O_{a}} peaks can be detected in the O 1s level (Figure \ref{fig:O1sN1sPt111}), even though the total pressure was divided by \num{6}, increasing the detection of photo-electrons.
    The peak at \qty{532}{\eV} is attributed to adsorbed water groups (\ce{H_2O_{a}}), reported between \qty{532.2} and \qty{532.9} eV on Pt(111) depending on the surface coverage \parencite{Fisher1980, Kiskinova1985}.

    \item[Ammonia:] Once oxygen is removed, gas phase water disappears from the O 1s level, confirming that water in gas phase is mostly present from the oxidation reaction rather than as UHV/gas contaminant.
    Adsorbed water is still visible from long desorption time, or from contaminants.
    The energy level differs by \qty{0.5}{\eV} from adsorbed water during reacting conditions, signifying different chemical environments.
    As observed in the mass spectrometer, the dissociation of ammonia towards nitrogen still occurs, a slightly shifted \ce{N_{2,g}} peak is reported at \qty{405.3}{\eV}.
    There is a clear effect of the presence of oxygen on the position of gas phase signals.
    The large peak linked to atomic nitrogen has disappeared, a peak linked to \ce{NH_{3,g}} is reported as well as a large and weak peak linked to \ce{NH_x} groups.

    \item[Argon:] The introduction of argon and removal of ammonia increases the \ce{H_2O_{g}} signal, possibly from contaminants. 
    Some nitrogen rich species are still visible. 
    Just like water, ammonia has the tendency to adsorb well on the inner walls of the chamber resulting in a slow pumping speed. 

    \item[Low oxygen:] Removing argon and introducing \qty{0.55}{\milli\bar} of oxygen has removed all the N 1s peaks by the oxidation of the leftover \ce{NH_x} species.
    As previously discussed, comparable peaks are observed in the O 1s level as under \qty{8.8}{\milli\bar} of oxygen, linked to hydroxyl groups, \ce{O_a}, and \ce{O_{2,g}} but with a higher apparent amount of atomic oxygen species in comparison to hydroxyl groups.

\end{description}

\subsubsection*{Pt 4f$_{7/2}$: Core level Evolution}

\begin{figure}[!htb]
    \centering
    \includegraphics[width=\textwidth]{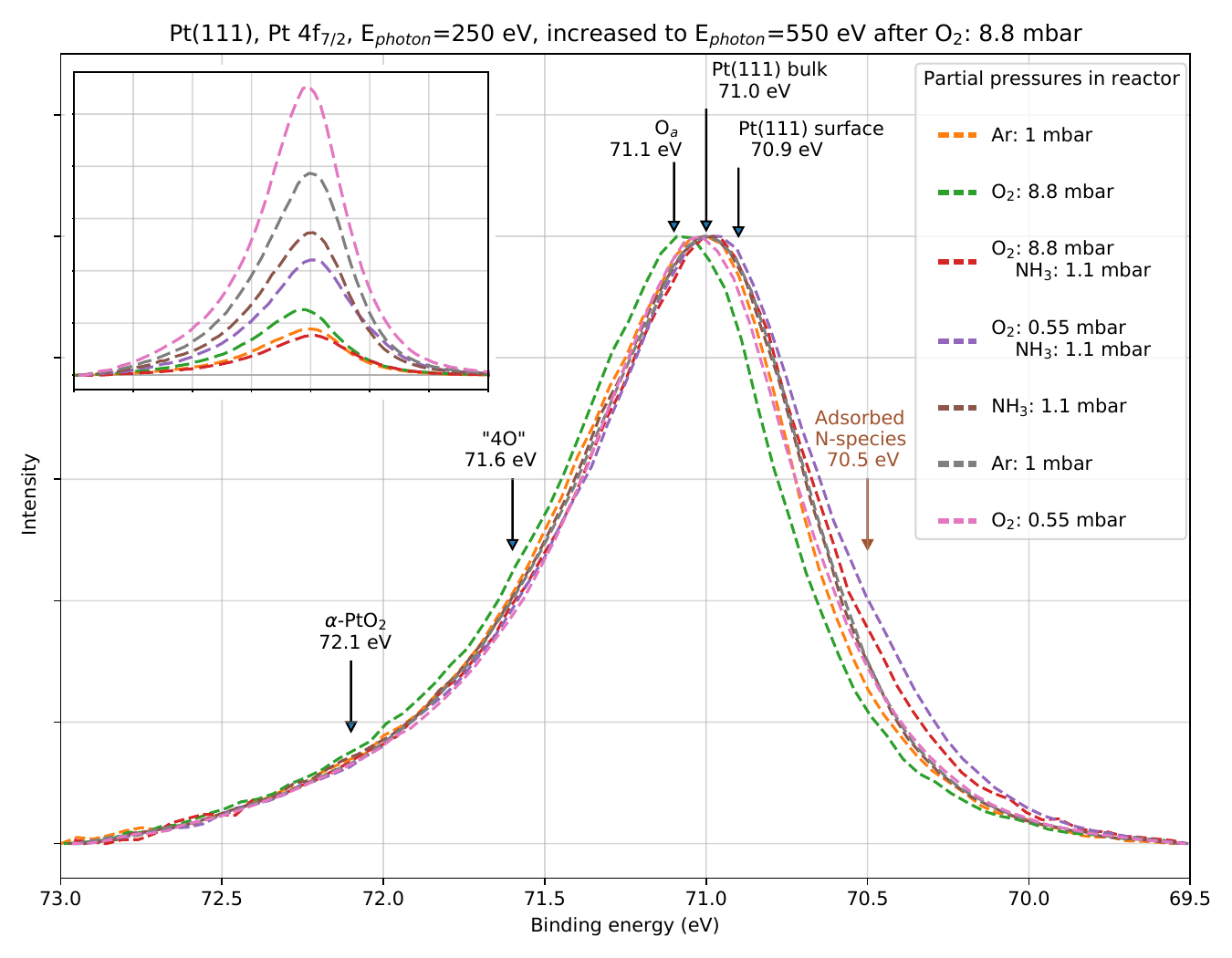}
    \caption{
        XPS spectra collected at the Pt 4f$_{7/2}$ level under different atmospheres at \qty{450}{\degreeCelsius} with an incoming photon energy of \qty{550}{\eV}.
        A Shirley-type background has been subtracted from all XPS spectra.
        Normalisation performed first by the background intensity and secondly by the maximum intensity to allow a qualitative comparison between different total pressures.
        The top‑left panel presents the spectra before scaling them to the maximum intensity.
    }
    \label{fig:Pt4fPt111}
\end{figure}

The Pt 4f$_{7/2}$ level was also measured, the spectra are normalised by the pre-edge (low binding energy side) intensity and shown in Figure \ref{fig:Pt4fPt111}.

\begin{description}
    \item[High oxygen:] Miller et al. \parencite*{Miller2011} have measured two peaks at \qtylist{72.1;73.5}{\eV} under \qty{6.66}{\milli\bar} of oxygen at \qty{450}{\degreeCelsius}.
    The \qty{72.1}{\eV} peak is assigned to (i) oxygen between the metallic surface and surface $\alpha$-\ce{PtO_2} oxide as reported by Ellinger et al. \parencite*{Ellinger2008}.
    The \qty{73.5}{\eV} peak is attributed to (ii) Pt atoms within the trilayer oxide structure.
    However, both peaks are absent at \qty{0.66}{\milli\bar} of oxygen at \qty{350}{\degreeCelsius}, for which the presence of (iii) ($2\times2$) chemisorbed oxygen and (iv) "4O" oxide surface stripes are linked to two other peaks, respectively \qty{71.1}{\eV}, and \num{71.6}-\qty{71.7}{\eV}.
    In a following study under \qty{13.33}{\milli\bar} of oxygen, slightly shifted high intensity peaks at \qty{72.2}{\eV} and \qty{73.6}{\eV} are linked to the presence of surface $\alpha$-\ce{PtO_2} oxide, fully covering a Pt(111) crystal \parencite{Miller2014}.
    The photon energy for both experiments is equal to \qty{275}{\eV}, whereas the photon energy is here equal to \qty{550}{\eV}.
    Thus, both studies by the Miller group are more sensitive to the surface structure since the emitted photo-electrons have increased inelastic mean free paths.
    Interestingly the crystal preparation to grow surface $\alpha$-\ce{PtO_2} oxide is not too far from this study.
    The Pt(111) crystal was exposed to \qty{13.33}{\milli\bar} of oxygen for \qty{10}{\minute} while cycling the temperature from \qtyrange{25}{525}{\degreeCelsius} four times.
    A small peak can be identified near \qty{71.6}{\eV} in the current experiment under \qty{8.88}{\milli\bar} of oxygen that could correspond to the "4O" oxide stripe structure.
    Its presence is not certain since the intensity is very low, corresponding potential peaks in the O 1s level cannot be resolved either.
    No clear peak could be detected near \qty{73.5}{\eV} in the Pt 4f$_{7/2}$ level, but a very small peak can be seen at \qty{72.1}{\eV}.
    Since the presence of surface $\alpha$-\ce{PtO_2} oxide was linked to very high intensity peaks by Miller et al. \parencite*{Miller2014}, this structure can safely be ruled out in this study.
    It seems that the temperature cycling is crucial to grow $\alpha$-\ce{PtO_2} on Pt(111).
    The peak maxima is shifted by \qty{0.09}{\eV} when introducing \qty{8.88}{\milli\bar} of oxygen in the cell after pure argon atmosphere.
    Overall, the lack of clear peak corresponding to the "4O" oxide stripe and $\alpha$-\ce{PtO_2} phases suggests that the oxygen on the Pt(111) surface is mostly chemisorbed, which could also be why the peak is slightly shifted towards \qty{71.1}{\eV}.
    
    \item[High oxygen reacting conditions:] When introducing \qty{1}{\milli\bar} of ammonia in the reactor, the maximum of the Pt 4f$_{7/2}$ peak is measured back at the position under inert atmosphere, while a new component is measured at \qty{70.5}{\eV}.
    
    \item[Low oxygen reacting conditions:] Reducing the pressure of oxygen to \qty{0.55}{\milli\bar} further increases the intensity of the latter component compared to the maximum peak intensity.

    \item[Ammonia:]Removing oxygen but keeping ammonia in the reactor removes this peak.
    Since the intensity of this peak increases when the \ce{O_2}/\ce{NH_3} ratio decreases, \textit{i.e.} when all the oxygen is consumed, it is probably linked to the presence of adsorbed nitrogen species on the platinum surface.
    The absence of this peak under the presence of ammonia in the cell, for which adsorbed nitrogen can not be detected, supports a link with \ce{N_a} (Figure \ref{fig:O1sN1sPt111}).
    
    \item[Argon:] Only a smaller shift is repeated when \qty{0.55}{\milli\bar} of oxygen is introduced after \qty{1}{\milli\bar} or argon, approximately equal to \qty{0.2}{\eV}.

    \item[Low oxygen:] Both spectra collected under argon before and after the oxidation cycle are very similar, some nitrogen species possibly left on the surface are probably removed by oxygen, which explains the increased intensity after the oxidation cycle near \qty{70.5}{\eV}.

\end{description}

\section{Discussion}

\subsection{Kinetics and transient oxides}

The different features observed in the SXRD maps during surface oxidation are summarised and discussed below.
First, two Pt(111)-($6\times6$)-R\ang{\pm 8.8} rotated hexagonal domains appear, at both high (Figure \ref{fig:MapsPt111A}b) and low (Figure \ref{fig:LargeMapsPt111LowOxygen}a) oxygen pressure.
Secondly, a Pt(111)-($8\times8$) superstructure grows, its second order peaks is detected before the first order peaks at both pressures (Figure \ref{fig:MapsPt111A}b-c \& Figure \ref{fig:LargeMapsPt111LowOxygen}).
More time resolved maps at low oxygen pressure allow us to be certain that this peak does not just appear during the map due to long measurement times (Figure \ref{fig:LargeMapsPt111LowOxygen}).
These maps also allow us to rule the disappearance of the grey and purple circled peaks in Figure \ref{fig:MapsPt111A}c as an alignment problems (large low intensity region), as they are clearly visible in Figure \ref{fig:LargeMapsPt111LowOxygen}.
X-ray reflectivity and SSR results indicate that the hexagonal structures are monolayers (Figure \ref{fig:Reflecto5} and Figure \ref{fig:LScans05}), while the Pt(111)-($8\times8$) superstructure eventually grows into a multi-layer thick epitaxied surface oxide (approximately \qty{15}{\angstrom} thick in Figure \ref{fig:LScansAndReflecto80}b).

\begin{figure}[!htb]
    \centering
    \includegraphics[width=\textwidth]{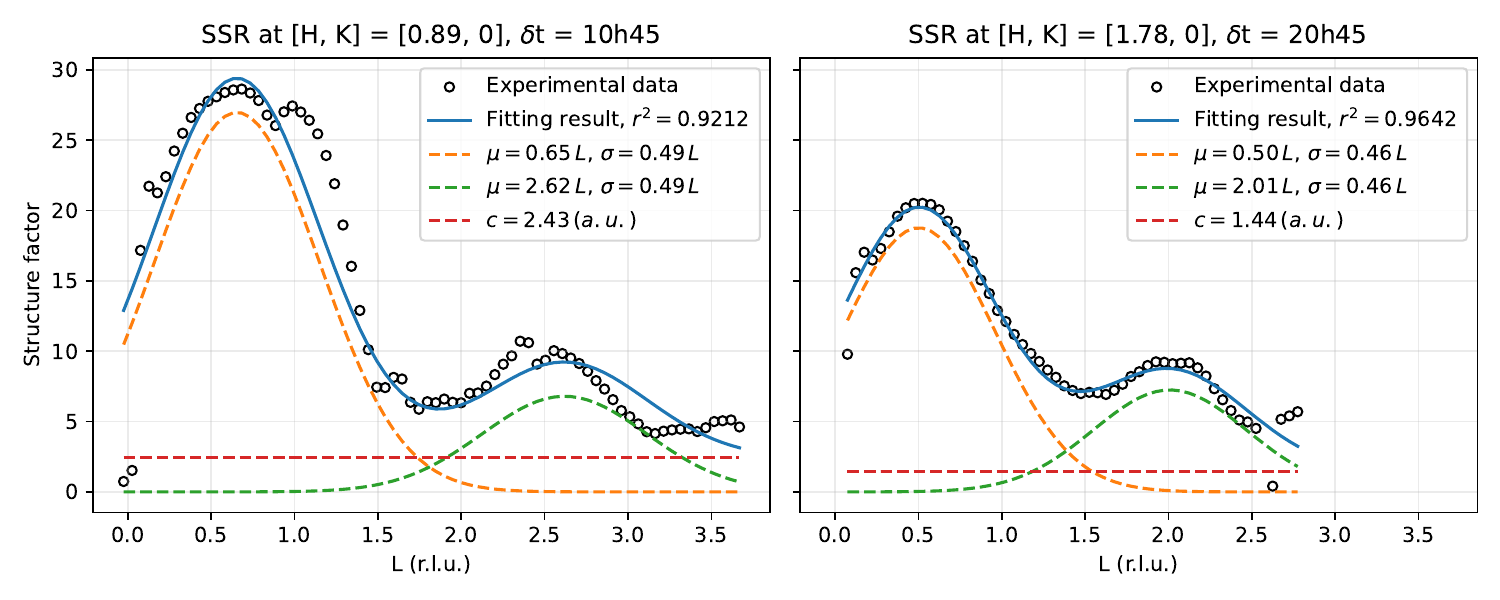}
    \caption{
        SSRs and fit results under \qty{80}{\milli\bar} of \ce{O2}.
        $\mu$, and $\sigma$ are the peak positions and standard deviations in $L$.
    }
    \label{fig:LScans80Fit}
\end{figure}

Seriani et al. \parencite*{Seriani2006} have determined that the most stable surface oxide on the Pt(111) surface is hexagonal $\alpha$-\ce{PtO_2}.
A high uncertainty regarding the lattice parameter in the $\vec{c}$ direction is reported in experimental and theoretically studies due to a poor crystallisation of the bulk oxide \parencite{Muller1968}, and to underestimation of weak interlayer Van der Waals forces in theoretical studies \parencite{Li2005}.
The formation of hexagonal on hexagonal surface $\alpha$-\ce{PtO_2} on Pt(111) is expected to induce a high amount of in-plane compressive strain in the oxide layer, from the large difference between the Pt-Pt distance on the (111) surface (\qty{2.77}{\angstrom}) and the in-plane lattice parameter of $\alpha$-\ce{PtO_2} (\qty{3.14}{\angstrom}).
A rotation of \ang{30} is expected to reduce the in-plane strain, resulting in a ($2\times2$) arrangement, no such structure was observed in this experiment on Pt(111).
Ellinger et al. \parencite*{Ellinger2008} have reported the existence of surface $\alpha$-\ce{PtO_2} on Pt(111) using SXRD at \qty{500}{\milli\bar} of oxygen, within \qty{245}{\degreeCelsius} and \qty{635}{\degreeCelsius}, and also with a Pt(111)-($8\times8$) superstructure, as previously reported by Ackermann \parencite*{Ackermann2007}.
The retrieved in-plane lattice parameter is equal to \qty{3.15}{\angstrom}, close to the value estimated by Seriani et al. \parencite*{Seriani2006}. 
The surface $\alpha$-\ce{PtO_2} model consists of one unit cell, with an out-of-plane lattice parameter reduced by \qty{15}{\percent} in comparison with the bulk structure (\qty{3.62}{\angstrom}) \parencite{Ellinger2008}.
They introduced a displacement of \qty{\approx33}{\percent} in the ($\vec{a}_{\alpha-\ce{PtO_2}}$ + $\vec{b}_{\alpha-\ce{PtO_2}}$) direction for the platinum atoms on the top layers with respect to those of the bottom layer.
A super-structure rod collected at [H, K] = [0, 0.88] present a peak at $L=0.65$ ($L$ computed using the distorted $\alpha$-\ce{PtO_2} unit cell).
The two SSRs performed in the current study were fitted using two Gaussian peaks sharing the same full width at half maximum (FWHM), and a constant background (Figure \ref{fig:LScans80Fit}).
The second peak near $L=1$ in the SSR at [H, K] = [0.89, 0] comes from a Pt powder signal, while the peak at $L=2.4$ is from alien signal.
A peak at [0.89, 0, 0.65] is visible, but using the Pt(111) surface unit cell, with an out-of-plane lattice parameter of \qty{6.79}{\angstrom}.
This is incompatible with the results presented in Ellinger et al \parencite*{Ellinger2008} since they used the lattice parameter of the refined $\alpha$-\ce{PtO_2} structure, $|\vec{c}|=\qty{4.16}{\angstrom}$.
It seems that a different out-of-plane structure is present in the current work.

XPS studies have also reported $\alpha$-\ce{PtO_2} above \qty{6.66}{\milli\bar} of \ce{O2} at \qty{450}{\degreeCelsius} \parencite{Miller2011, Miller2014}.
As mentioned by Fantauzzi et al \parencite*{Fantauzzi2017}, the extended exposure to oxygen necessary to grow the oxide structure makes its characterisation difficult, the structure possibly evolving with time.
A large peak at \qty{530.7}{\eV} was attributed to Pt surface oxides during a study at $p_{O2} = \qty{1}{\bar}$ \parencite{VanSpronsen2017}.
Additional simulations are needed to fully understand the structures in this study using the proposed distorted $\alpha$-\ce{PtO_2} structure, but also different surface models including the \textit{spoked-wheel} superstructure identified \textit{via} scanning tunnelling microscopy (STM) by Van Spronsen et al \parencite*{VanSpronsen2017} and Boden et al. \parencite*{Boden2022}.
This structure, identified above \qty{1}{\bar} of oxygen, and at approximately \qty{200}{\degreeCelsius}, also exhibits a ($8\times8$) coincidence with the Pt(111) lattice.
    
\begin{figure}[!htb]
    \centering
    \includegraphics[width=\textwidth]{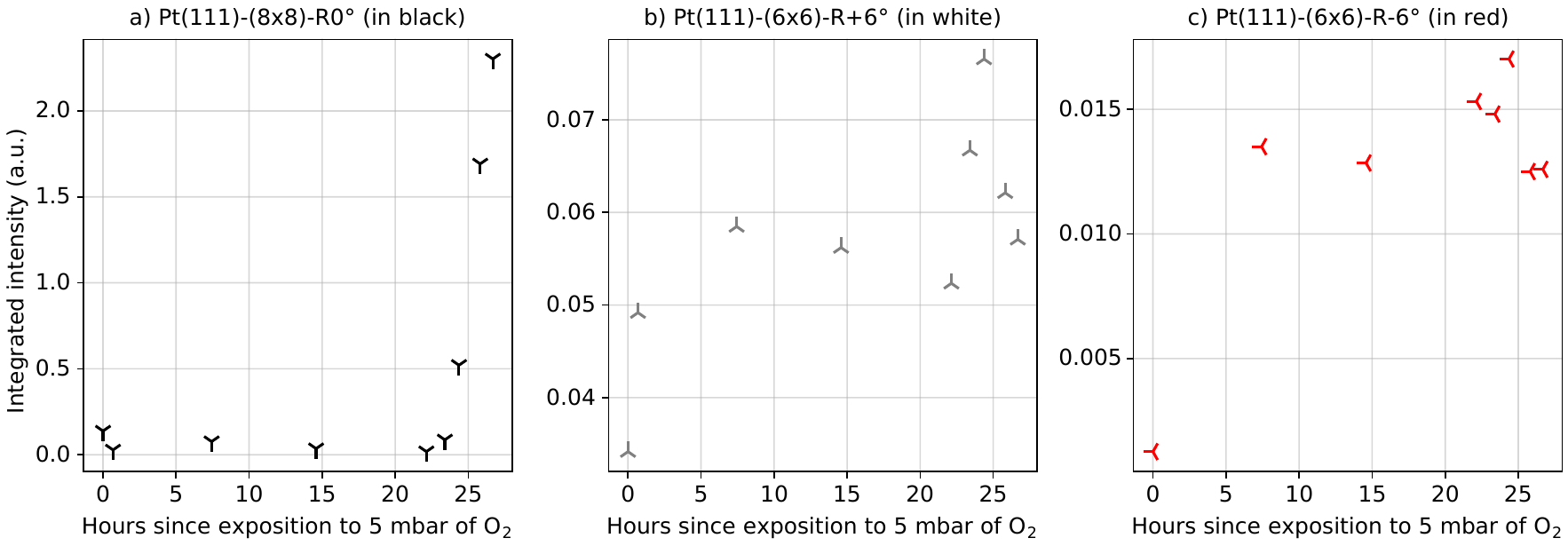}
    \caption{
        Intensity evolution as a function of time since introduction of \qty{5}{\milli\bar} of oxygen for peaks corresponding to the Pt(111)-($6\times6$)-R\ang{\pm8.8} and Pt(111)-($8\times8$) structures.
    }
    \label{fig:HexBraggPeaks}
\end{figure}

A more quantitative analysis of the different structures appearing during the exposition to oxygen was performed by integrating the scattered intensity around the Pt(111)-($8\times8$) and two Pt(111)-($6\times6$)-R\ang{\pm8.8} signals present in the (100) region (respectively circled in black, red and white in Figure \ref{fig:SmallMapsPt111LowOxygen}).
The average background was subtracted to each reciprocal space voxel before integration.
The starting time of each reciprocal space in-plane map is used as estimate for the time since the introduction of oxygen, the evolution of each peak is presented in Figure \ref{fig:HexBraggPeaks}.
The growth of the Pt(111)-($8\times8$) peak at [H, K] = [0, 0.89] follows an exponential increase after \qty{23}{\hour} of exposition.

The white Pt(111)-($6\times6$)-R\ang{\pm8.8} peak is visible from the start of the exposition to oxygen (Figure \ref{fig:SmallMapsPt111LowOxygen}a), whereas the red Pt(111)-($6\times6$)-R\ang{\pm8.8} peak is only clearly visible after \qty{7}{\hour} (Figure \ref{fig:SmallMapsPt111LowOxygen}c), with a lower intensity.
Both peaks intensity quickly increases in the first hours of oxygen exposure, and then reach a steady state between \qtylist{8;23}{\hour}, increasing again together with the appearance of the Pt(111)-($8\times8$) signal.
The Pt(111)-($6\times6$)-R\ang{\pm8.8} peak intensity then starts to decreases after \qty{\approx 25}{\hour}.
Additional studies are needed to understand if whether or not the rotated hexagonal structures act only as precursors for the Pt(111)-($8\times8$) superstructure, or if those peaks are still present after an extended exposure to oxygen.
Several SSRs have been measured at the positions of the Pt(111)-($6\times6$)-R\ang{\pm8.8} peaks at low oxygen pressure (\qty{5}{\milli\bar}), presented in Figure \ref{fig:LScans05}.
The intensity is constant as a function of $L$, similarly to the SSRs presented in Figure \ref{fig:LScansAndReflecto80}, which shows that the rotated hexagonal structures are flat monolayers on the Pt(111) surface.
No SSR was measured at the position of the Pt(111)-($8\times8$) peak at that condition.

\subsection{Oxides and ammonia}

The presence of $\alpha$-PtO$_2$ on Pt(111) has been shown to not hinder the oxidation of \ce{CO} by Ackermann et al. \parencite*{Ackermann2007}, that would then occur \textit{via} a Mars-Van Krevelen mechanism \parencite{Mars1954}, inducing a progressive roughening of the platinum surface.
The presence of $\alpha$-\ce{PtO_2} on the Pt(111) surface was also shown to provide favourable special sites that could contribute to the catalytic activity \parencite{Li2005}.
The low roughness and removal of SXRD signals during ammonia oxidation is in contradiction with a possible Mars-Van Krevelen mechanism involving surface oxides at the current conditions, consistent with the reported Langmuir-Hinshelwood literature mechanism.

\begin{figure}[!htb]
    \centering
    \includegraphics[width=\textwidth]{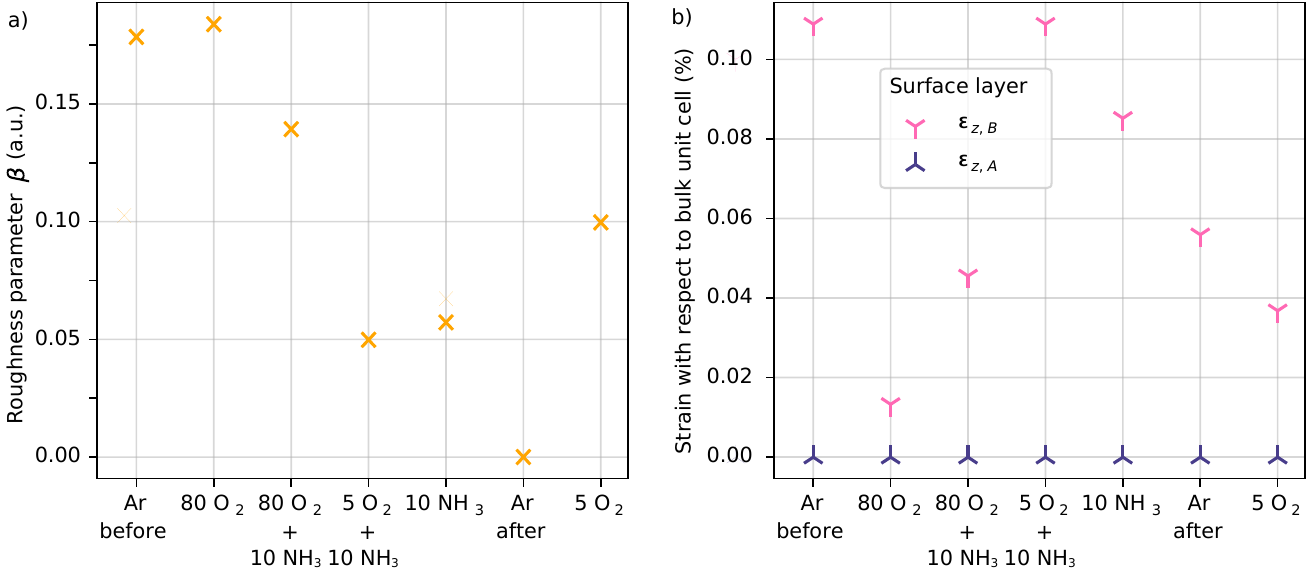}
    \caption{
        a) Fitting results for roughness parameter $\beta$ and b) out-of-plane strain $\epsilon_z$ with respect to bulk lattice parameter, as a function of the experimental conditions.
        The innermost layer is named $A$, which has the same lattice parameter as in the bulk.
        The topmost layer of the Pt(111) single crystal is named $B$. 
    }
    \label{fig:CTRFit111}
\end{figure}

The presence of adsorbed water supports a Langmuir-Hinshelwood mechanism with quick stripping of hydrogen from \ce{NH_{x,a}} species by \ce{OH_a} and \ce{O_a}, eventually forming adsorbed water.
The de-hydrogenation process must be limiting the catalytic activity due to the lack of available adsorbed oxygen species near adsorbed ammonia, which could be why \ce{NH_{3,a}} is observed, but not \ce{NH_{x,a}}.
The key importance of available atomic oxygen for the production of \ce{NO} is supported by the simultaneous absence of \ce{O_a} in the O 1s level, and the absence of \ce{NO} in the RGA signals (Figure \ref{fig:XPS111RGA}).
The presence of adsorbed nitrogen could be due to long residual times on the catalyst before recombination and desorption of \ce{N_2}.

The dissociation of ammonia is reported to be slower without the help of oxygen species on Pt(111) \parencite{Offermans2006,Offermans2007, Imbihl2007, NovellLeruth2008}.
This could explain why such a large peak is observed for \ce{NH_{x,a}}, and why \ce{NH_{3,g}} is observed rather than \ce{NH_{3,a}}, since most of the adsorption sites are probably occupied by \ce{NH_{x,a}} species.
The difficulty to fully dissociate ammonia is also correlated to the production of hydrogen, it is possible that the combination of two hydrogen atoms removed from ammonia to produce \ce{H_2} is slow, and thus occupies part of the adsorption sites in the absence of oxygen.

To investigate potential surface relaxation effects, the three CTRs were fitted together (see Methods), the evolution of the surface roughness and out-of-plane strain is shown in Figure \ref{fig:CTRFit111}.
The Pt(111) surface roughness is high under Argon at \qty{450}{\degreeCelsius} (Figure \ref{fig:CTRFit111}a), probably due to the presence of impurities contained in the argon gas flow.
The introduction of oxygen in the cell further increases the surface roughness, as expected from the formation of the different surface oxides visible in the in-plane reciprocal space maps (Figure \ref{fig:MapsPt111A}, CTR recorded between c and d).
Adding ammonia in the reactor cell, which removes the different surface oxides (Figure \ref{fig:MapsPt111B}), also decreases the surface roughness.
Finally, the re-introduction of \qty{5}{\milli\bar} of oxygen increases the surface roughness again (Figure \ref{fig:CTRFit111}a), in accordance with the formation of surface oxides detected during in-plane reciprocal space maps (Figure \ref{fig:LargeMapsPt111LowOxygen}).
Low amounts of strain are detected on the surface, almost imperceptible when observing the CTR minimum position in Figure \ref{fig:CTRPt111}.
From the fitting results, the topmost layer is already under tensile strain under inert atmosphere.
Different reacting conditions are related to the same direction of out-of-plane displacement, higher ammonia to oxygen ratio coincides with higher tensile strain, no changes from tensile to compressive strain are observed during the reaction.
The largest evolution in the strain values comes from the introduction oxygen after pure argon, the out-of-plane lattice parameter almost returning to its bulk value.
Overall, the presence of oxygen leads to the formation of surface oxides which lowers the surface strain, and increases the surface roughness.
Reacting conditions and the presence of ammonia remove the surface oxides, and decrease the surface roughness.

\section{Conclusion}

Although Pt--Rh alloys and Pt(100) surfaces can undergo reversible oxide formation and reconstructions that affect ammonia oxidation \cite{Resta2020a, Simonne2026a}, the \textit{operando} surface state of pure Pt(111) has been less well established under realistic reaction environments. 
Here we combined surface X-ray diffraction (SXRD), crystal truncation rod (CTR) analysis, and near-ambient pressure X-ray photoelectron spectroscopy (NAP-XPS) with on-line mass spectrometry to directly link surface structure, chemical state, and product distribution during ammonia oxidation on Pt(111). 
Across the conditions explored, Pt(111) does not sustain a stable surface oxide or reconstruction in the presence of ammonia. 
Instead, oxygen-rich feeds promote transient ordered phases, including hexagonal monolayers and a Pt(111)-($8\times8$) superstructure, which are removed upon ammonia exposure.
The combined NAP-XPS and product analysis are consistent with a reaction regime governed by co-adsorbed \ce{NH_x} and oxygen-derived species, with oxygen-rich conditions favouring \ce{NO} production and ammonia-rich feeds shifting selectivity toward \ce{N2}.
The availability of adsorbed atomic oxygen is clearly revealed as the key to NO selectivity on Pt(111).
This behaviour contrasts with Pt(100), where epitaxial oxide phases such as \ce{Pt3O4}(001) can form at similar oxidation condition, and where difference surface reconstructions drive the catalyst selectivity \cite{Simonne2026a}. 
These results highlight a pronounced facet dependence in the stability of oxidized surface states on platinum and demonstrate that Pt(111) operates predominantly in an oxide-free regime under ammonia oxidation.
More broadly, this work frames ammonia oxidation as an interface problem in which dynamic surface structure and surface chemistry jointly determine function. 
By resolving ordered surface phases, relaxations, and roughness under near-industrial environments, SXRD provides a uniquely powerful route to establish structure--chemistry relationships at reactive metal/oxide interfaces. 
The present findings therefore provide experimentally grounded guidance for designing platinum-based catalytic interfaces with improved stability and selectivity under strongly oxidizing operating conditions.

\section{Materials and methods}

\subsection{Surface X-ray Diffraction}

The Pt(111) sample was bought from Surface Preparation Laboratory \cite{SPL}, polished to mirror-finish.
It is circular, \qty{1}{\cm} large, and \qty{0.2}{\cm} thick (Figure \ref{fig:SampleSXRD}b). 
The polished surface area of the single crystal is \qty{\approx 50}{\mm^2}.
The surface X-ray diffraction experiment was performed at the SixS beamline (SOLEIL synchrotron), at \qty{18.44}{\keV} on the vertical Z-axis diffractometer by means of a flow reactor that operates mirroring ref. \cite{VanRijn2010} (Figure \ref{fig:SampleSXRD} \& \ref{fig:SampleHolder}) at industry relevant conditions.
The large reactor chamber accommodates a sputtering gun, and is connected to two gas inlet and outlet valves.
The incoming gas mixture is controlled upstream by mass flow controllers, while the total reactor pressure is set at the outlet.
A leak from the reactor chamber to a residual gas analyser (RGA) allows monitoring the reactant and product partial pressures.
Two sputtering (\qty{30}{\min}) and annealing (\qty{800}{\degreeCelsius} - \qty{5}{\min}) cycles were necessary to obtain a clean and flat surface before starting the SXRD experiment, as verified by the truncated Pt(111) peaks in Figure \ref{fig:MapsPt111A}.
The sputtering current was maximised by tuning the sample height to ensure that the Ar ion beam covers the entire sample surface.
The incident angle during the SXRD experiment was set to \ang{0.3} to increase surface sensitivity.
The hexagonal basis set of vectors with $|a_{1}|$, $|a_{2}|$ lying in the surface plane and $|a_{3}|$ perpendicular to them was used to describe the crystal structure. 
In terms of the bulk lattice constant $a_{Pt}$ ($a_{Pt}$ = \qty{3.92}{\angstrom}) the lengths of these vectors can be expressed as $|a_{1}|$=$|a_{2}|$= $|a_{Pt}| / \sqrt{2} $= \qty{2.77}{\angstrom} and $|a_{3}|$ = $\sqrt{3}a_{Pt}$ = \qty{6.79}{\angstrom}.
Given the Pt(111) surface three-fold symmetry and for the sake of temporal optimisation, only a third of the reciprocal space in the ($\vec{q}_x$, $\vec{q}_y$) plane, where $\vec{q}_x$ and $\vec{q}_y$ are the in-plane components of the scattering vector, was measured during the in-plane reciprocal space maps.
The (2$\bar{1}$L), ($10L$) and ($11L$) crystal truncation rods were measured \qty{6}{\hour} after the start of each condition, each measurement lasted for \qty{2}{\hour}.
The rods where numerically integrated using the \textit{fitaid} module of \textit{BINoculars} which allows for background corrections \parencite{Roobol2015}.

\subsection{X-ray reflectivity}

The specular X-ray reflectivity simulations are conducted using the Parratt algorithm \parencite{Parratt1954}.
The classical way of implementing roughness is based on a Gaussian distribution, introducing corrective factors to the electric field amplitudes at the interfaces in accordance with the Nevot-Croce model \parencite{Nevot1980}.
For the computation of material refractive indices, the Henke tables \parencite{Henke1993} are employed.
A simple model was used consisting of a semi-infinite slab of platinum, on top of which is present a homogeneous layer of platinum oxide.
The density of the layer ($\rho_{ox}$) was set free during the minimisation process, between zero and \qty{21.2e-3}{FU\per\cubic\angstrom} (formula unit per cubic Angström, computed from unit cell elements and volume), the value for $\alpha$-PtO$_2$.
The initial intensity is fixed to obtain the best match at low incident angles, the intensity drop after the critical angle is difficult to fit, but the period of the oscillations, as well as the intensity at high angles are in general well represented.
The oscillation magnitude can be linked to the density of the oxide layer on top of the Pt(111) surface.
The fitted values (\qty{7.e-3}{FU\per\cubic\angstrom} and \qty{6.5e-3}{FU\per\cubic\angstrom}) are an order of magnitude below the expected value for a homogeneous layer of $\alpha$-PtO$_2$.
The root mean square roughness of the oxide layer $\sigma_{ox}$ is always zero, whereas the roughness of the Pt(111) surface ($\sigma_{sub}$) is of the same order of magnitude for both measurements.

\subsection{Near Ambient Pressure X-ray Photoelectron Spectroscopy}

The same Pt(111) crystal was measured with NAP-XPS at the B07 beamline\cite*{Held2020} (Diamond Light Source).
Due to the impact of the pressure on the electron transmission, the total pressure was lowered and no longer kept constant by the use of a carrier gas (argon) between each condition. 
The reactant partial pressure becomes approximately \qty{10}{\percent} of the partial pressures used in SXRD, with an equivalent (\ce{O2}/\ce{NH3}) ratio (Table \ref{tab:Combined_Conditions_SXRD_XPS}).
The electron analyser axis was set at normal emission with respect to the sample surface.
The cone of the analyser was also used as inlet for mass spectrometry measurements, it has a diameter of \qty{0.3}{\mm}, with a typical working distance of \qty{0.2}{\mm}.
The combined energy resolution of the beamline and analyser, while working in NAP conditions, ranged between \qty{0.8}{\eV} and \qty{1.2}{\eV}.
The Fermi edge was also measured after each core level to be able to correctly calibrate the binding energy.

\subsection*{Acknowledgements}

We would like to offer our most sincere gratitude to Dr. Michele Sauvage for fruitful discussions, and to Dr. Fréderic Picca for his help regarding the installation and development of \textit{BINoculars} (https://repo.or.cz/hkl.git, https://github.com/picca/binoculars/) at SixS.
This project has received funding from the European Research Council (ERC) under the European Union’s Horizon 2020 research and innovation program (grant agreement No. 818823).
We thank SOLEIL for providing the majority of the beam time required for this work and Diamond Light Source for the additional allocation, and we are grateful to the SixS and B07 teams for their assistance throughout the experiments.

\subsection*{Conflict of interest}

The authors declare no competing financial interest.

\appendix

\renewcommand\thefigure{\thesection.\arabic{figure}}    
\setcounter{figure}{0}    

\section{Reactor calibration}

\begin{figure}[!htb]
    \centering
    \includegraphics[width=0.495\textwidth]{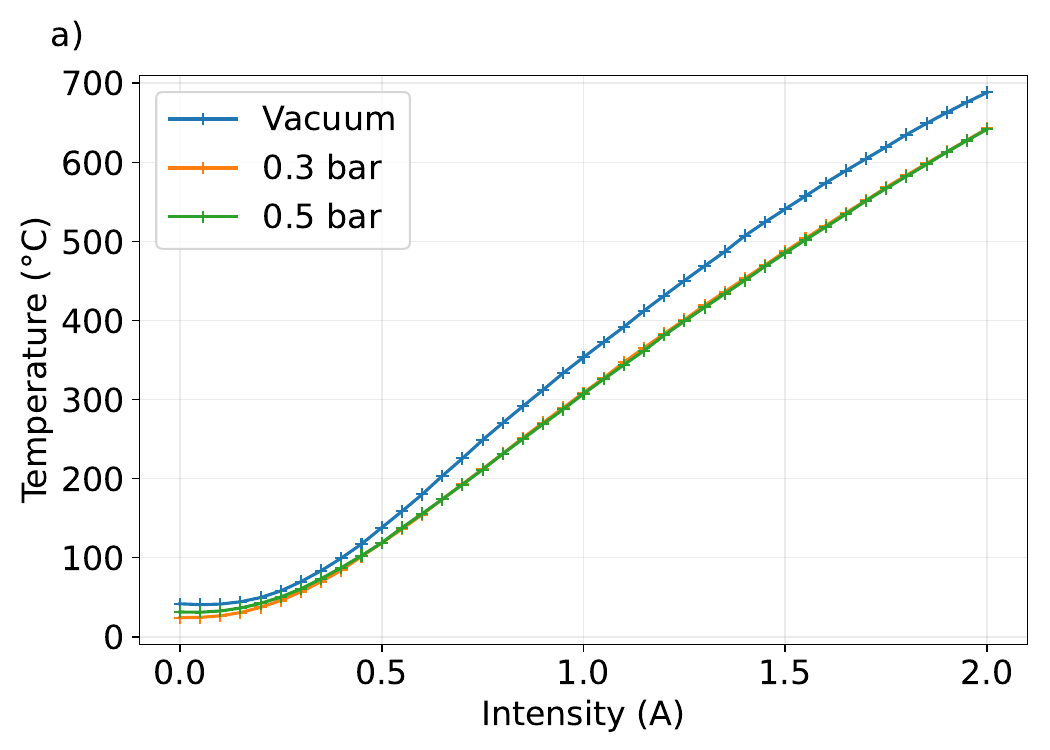}
    \includegraphics[width=0.495\textwidth]{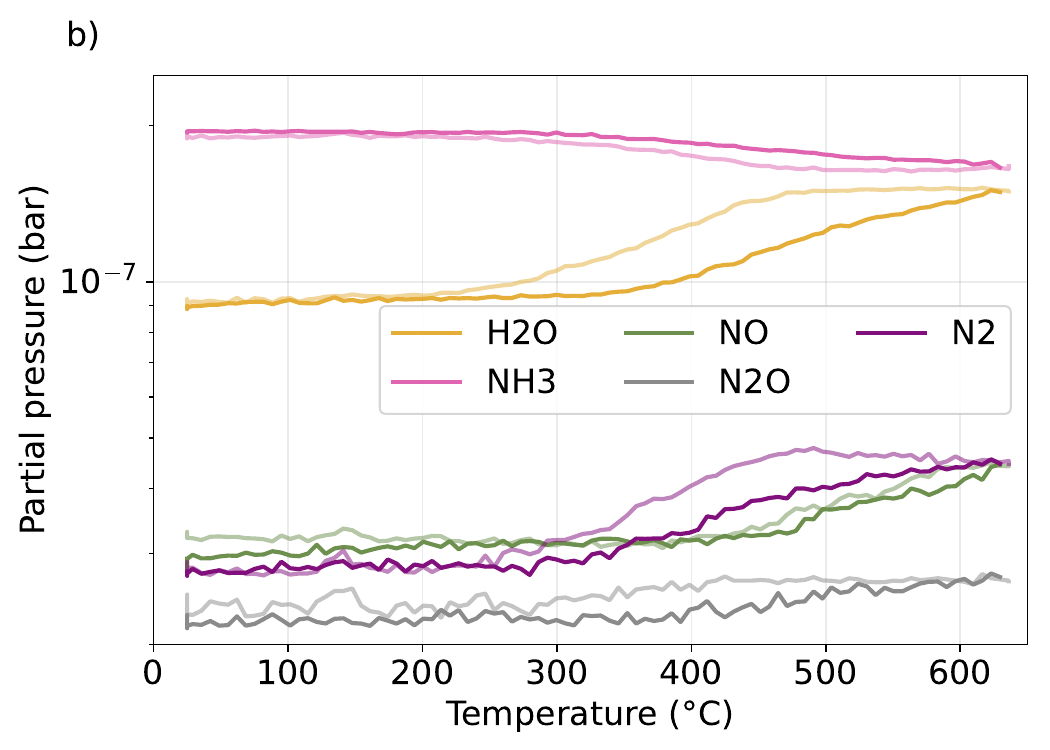}
    \caption{
        a) Sample temperature calibration, measured under vacuum and different \ce{Ar} pressures as a function of heater input current.
        b) Evolution of partial pressures in the RGA chamber when $p_{O_2}/p_{NH_3}=8$.
        Increasing and decreasing (low transparency) temperature ramp to \qty{650}{\degreeCelsius}.
        The ramp is performed with 100 steps, each lasting \qty{10}{\second}.
    }
    \label{fig:TempRamps}
\end{figure}

\begin{table}[!htb]
\centering
    \begin{tabular}{@{}llllllll@{}}
    \toprule
                       & \ce{Ar}    & \ce{NH3}  & \ce{O2}   & \ce{NO}    & \ce{N2O}  & \ce{N2}   & \ce{H2O} \\
    \midrule
    \qty{25}{\degreeCelsius} & \num{17.7} & \num{25.1} & \num{26.5} & \num{25.9} & \num{17.4} & \num{26.0} & \num{18.6} \\
    \qty{125}{\degreeCelsius} & \num{22.4} & \num{37.2} & \num{34.0} & \num{33.1} & \num{26.0} & \num{32.8} & \num{26.1} \\
    \qty{225}{\degreeCelsius} & \num{26.5} & \num{53.1} & \num{41.0} & \num{39.6} & \num{34.1} & \num{39.0} & \num{35.6} \\
    \qty{325}{\degreeCelsius} & \num{30.3} & \num{68.6} & \num{47.7} & \num{46.2} & \num{41.8} & \num{44.8} & \num{46.2} \\
    \bottomrule
    \end{tabular}%
\caption{Thermal conductivity of gases used in during ammonia oxidation in \unit{\mW \per \meter \per \kelvin} \parencite{ThermalConductivityOfGases}.}
\label{tab:ThermalConductivity}
\end{table}
\clearpage

\section{Surface X-ray diffraction}

\begin{table}[!htb]
    \centering
    \resizebox{\textwidth}{!}{%
    \begin{tabular}{@{}|l|l|lllllllllll|@{}}
        \toprule
        Structure & Interplanar & \multicolumn{11}{c|}{Oxygen pressure} \\
        \midrule
          & spacing (\unit{\angstrom}) & \multicolumn{2}{l|}{80 mbar} & \multicolumn{9}{l|}{5 mbar} \\
        \midrule
         & & \multicolumn{11}{c|}{Time since gas introduction (end of measurement)} \\
        \midrule
                                     &                  & 03h23 & 09h53 & \multicolumn{1}{|l}{00h34} & 04h03 & 08h00 & 15h57 & 22h56 & 24h08 & 25h43 & 26h36 & 27h28 \\
        \midrule 
        Pt(111)-($6\times6$)-R\ang{\pm 8.8} & $3.01 \pm 0.01$ & \yes  & \yes  & \multicolumn{1}{|l}{\yes}  & \yes  & \yes  & \yes  & \yes  & \yes  &                              \yes & \yes & \yes \\
        \midrule 
        Pt(111)-($6\times6$)-R\ang{\pm 8.8} & $2.92 \pm 0.03$ & \yes  & \yes  & \multicolumn{1}{|l}{\no}   & \no   & \no   & \no   & \no   & \no   &                              \yes & \no & \no \\
        \midrule 
        Pt(111)-($6\times6$)-R\ang{\pm 8.8} & $2.87 \pm 0.02$ & \no   & \no   & \multicolumn{1}{|l}{\yes}  & \yes  & \yes  & \yes  & \yes  & \yes  &                              \yes & \yes & \yes \\
        \midrule 
        Pt(111)-($6\times6$)-R\ang{\pm 8.8} & $2.79 \pm 0.07$ & \no   & \no   & \multicolumn{1}{|l}{\no}   & \no   & \yes  & \yes  & \yes  & \yes  &                              \yes & \yes & \yes \\
        \midrule 
        Pt(111)-($8\times8$)       & $2.69 \pm 0.02$ & \no   & \yes  & \multicolumn{1}{|l}{\no}   & \no   & \no   & \no   & \no   & \yes  &                              \yes & \yes & \yes \\
        \midrule 
        Pt(111)-($8\times8$)       & $1.53 \pm 0.00$ & \yes & nv     & \multicolumn{1}{|l}{nv}    & \yes  & nv    & nv    & nv & nv & nv & nv & nv \\
        \bottomrule
    \end{tabular}
    }
    \caption{
        Interplanar spacings computed from signals observed during in-plane reciprocal maps, for different oxygen pressure and exposition times for Pt(111).
        The three markers (\yes, \no, \text{nv}) correspond to observed, non-observed, and non visible signals (\textit{i.e.} not in the area spanned by the map).
        The errors on the interplanar spacings are computed by considering the positions of similar peaks in q-space.
    }
    \label{tab:InterplanarSpacingsPt111Oxygen}
\end{table}

\begin{figure}[!htb]
    \centering
    \includegraphics[width=\textwidth]{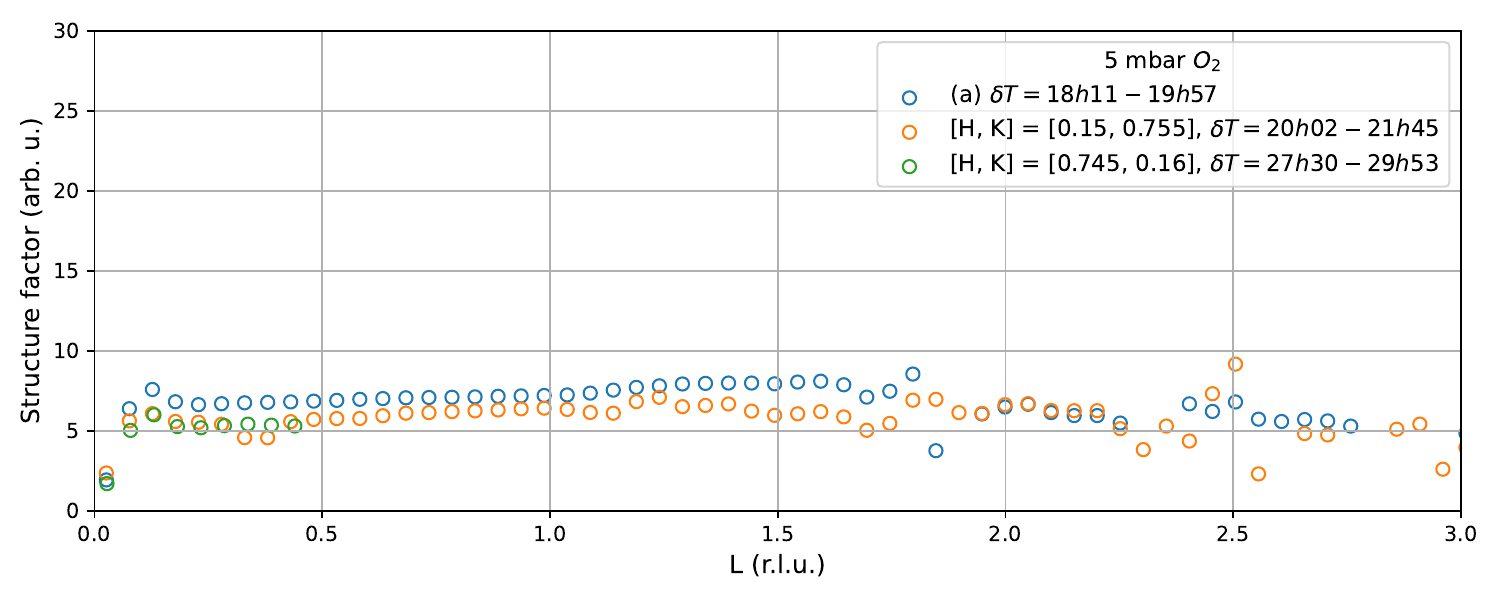}
    \caption{
        SSR measured at $p_{O_2} = \qty{5}{\milli\bar}$.
    }
    \label{fig:LScans05}
\end{figure}

\begin{figure}
    \centering
    \includegraphics[width=\textwidth]{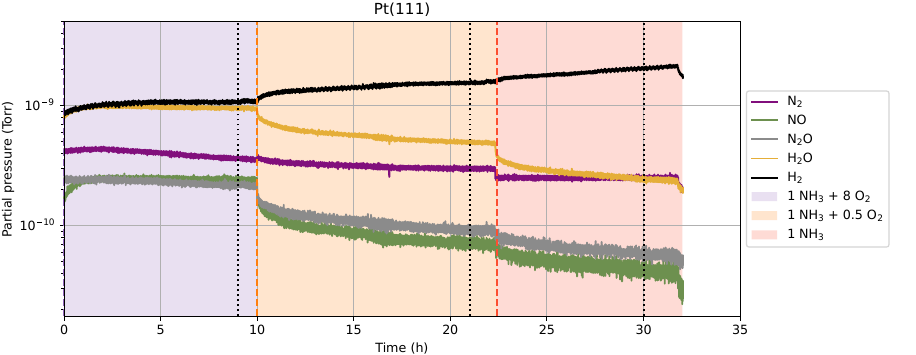}
    \caption{Time dependent partial pressures recorded from a leak in the reactor output by a residual gas analyser (RGA) during the SXRD experiment on the Pt(111) single crystal at \qty{450}{\degreeCelsius}.}
    \label{fig:RGA450Pt111}
\end{figure}

\clearpage 

\section{X-ray Photoelectron Spectroscopy}

\begin{figure}[!htb]
    \centering
    \includegraphics[width=\textwidth]{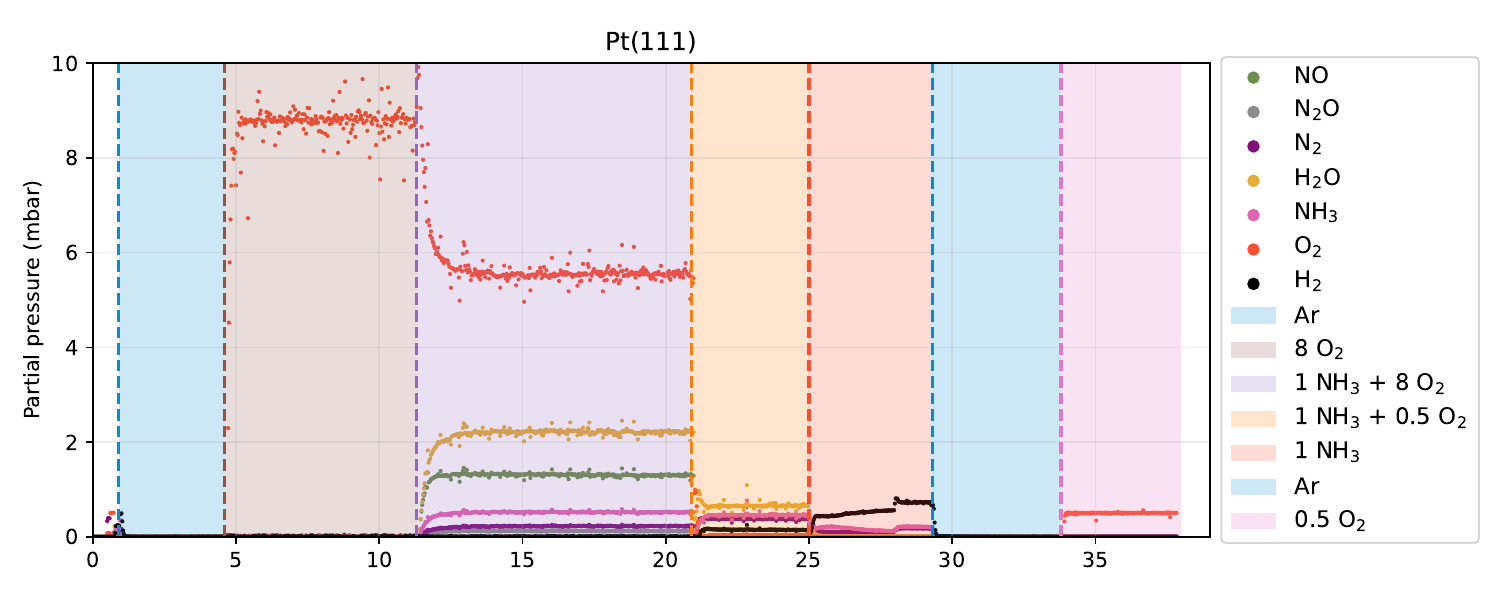}
    \caption{
        Evolution of reaction product partial pressures as a function of time during the XPS experiment on Pt(111) at \qty{450}{\degreeCelsius}.
        Transition between conditions are indicated with dashed vertical lines.
        The glitch in the partial pressure at \qty{28.5}{\hour} is due to a re-setting of the total pressure to \qty{1.1}{\milli\bar}.
    }
    \label{fig:XPS111RGA}
\end{figure}

\begin{figure}[htb!]
    \centering
    \includegraphics[width=\linewidth]{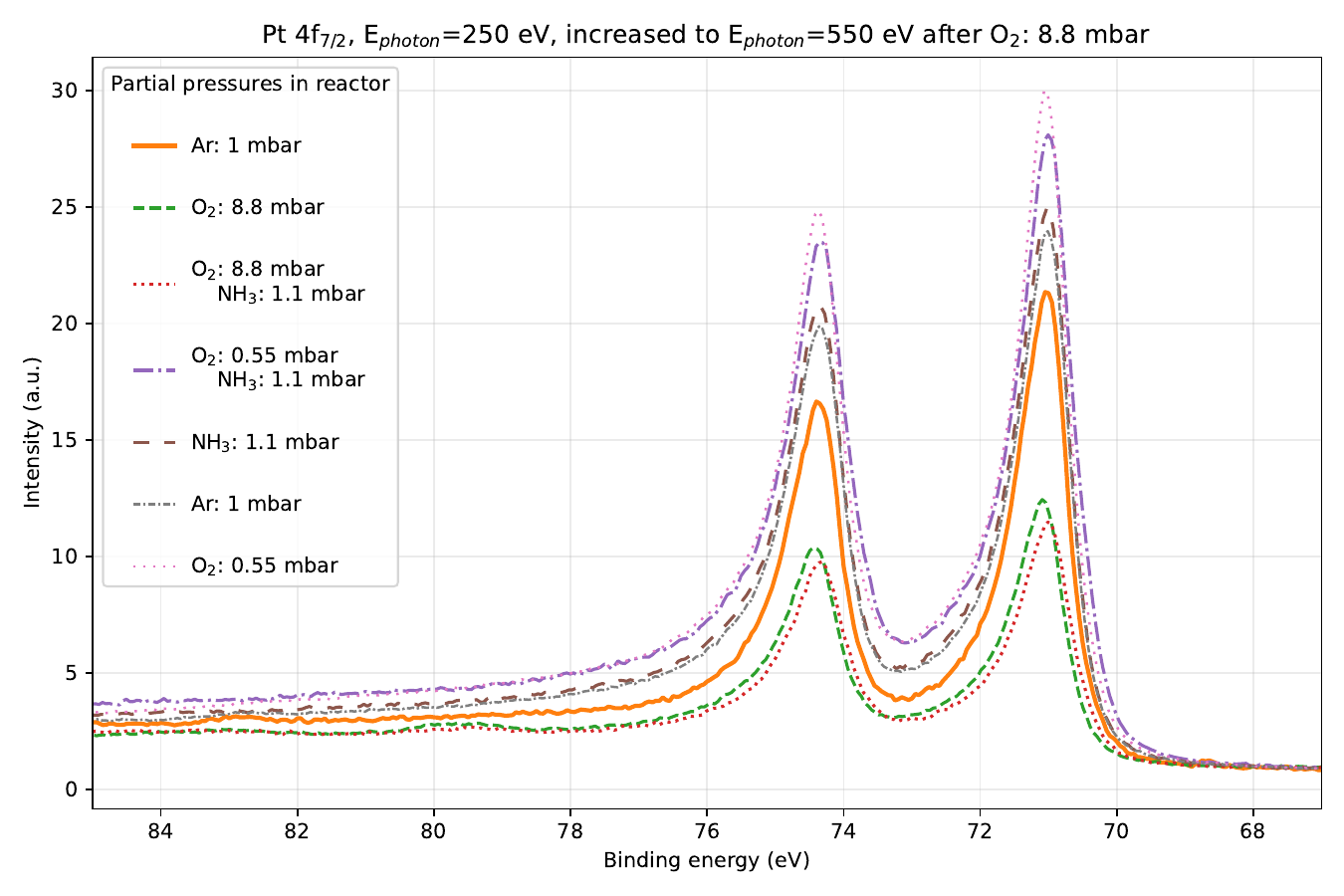}
    \caption{XPS spectra collected at the Pt 4f$_{7/2}$ level under different atmospheres at \qty{450}{\degreeCelsius}. Normalisation performed by the pre-edge intensity.
    }
    \label{fig:Pt4fExtendedRange}
\end{figure}

\clearpage 

\section{Experimental setups}

\begin{figure}[!htb]
    \centering
    \includegraphics[width=\textwidth]{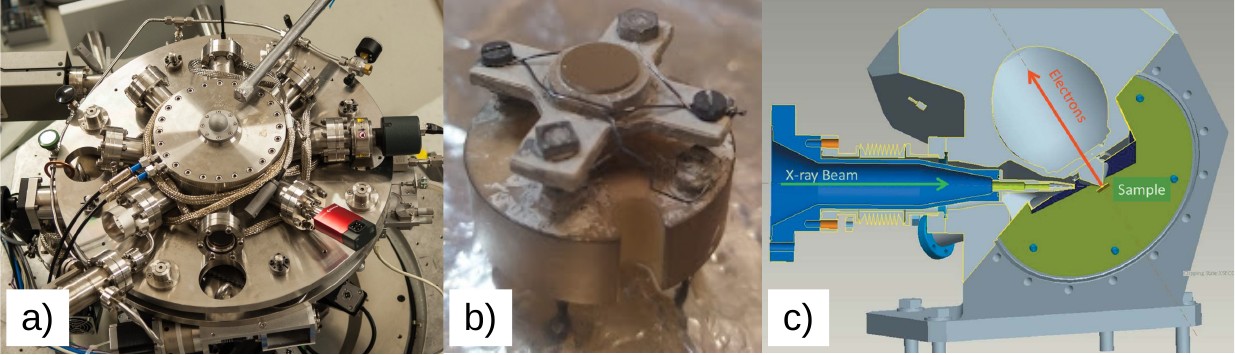}
    \caption{
        a) Large XCAT reactor.
        b) Round single crystal set on sample holder, the crystal presents here a (111) top surface.
        The surface area of the single crystal is \qty{\approx 50}{\mm^2}.
        c) Detailed drawing of the XPS chamber at the B07 beamline at the Diamond Light Source \parencite{Held2020}.
    }
    \label{fig:SampleSXRD}
\end{figure}

\begin{figure}[!htb]
    \centering
    \includegraphics[width=0.7\textwidth]{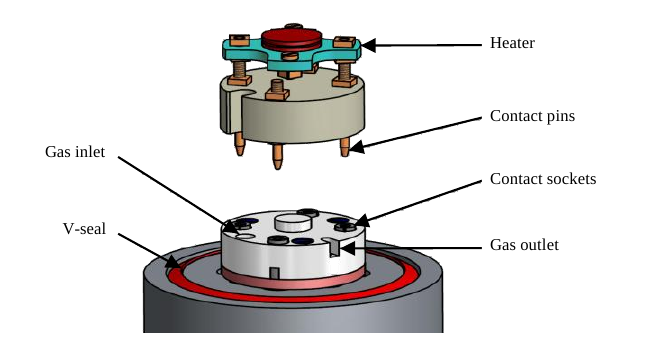}
    \caption{
        Sample holder used for the XCAT catalysis reactor. A PEEK dome, transparent to X rays, is set upon the sample holder during operando measurements. 
    }
    \label{fig:SampleHolder}
\end{figure}

\clearpage

\printbibliography

\end{document}